\documentclass[11pt]{revtex4}
\usepackage{mathtools}
\usepackage{hyperref}
\usepackage{amsmath,amsfonts}
\usepackage{amssymb}
\usepackage{ulem}
\allowdisplaybreaks

\begin{document}

\begin{flushright} {\footnotesize YITP-26-08, RESCEU-2/26, IPMU26-0002}  \end{flushright}

\title{Propagation and polarization of gravitational waves on curved spacetime backgrounds in Einstein-\AE ther theory}

\author{Yu-Qi Dong$^{a,b,c}$}
\email{dongyq2023@lzu.edu.cn (corresponding author)}

\author{Shinji Mukohyama$^{c,d,e}$}
\email{shinji.mukohyama@yukawa.kyoto-u.ac.jp}

\author{Yu-Xiao Liu$^{a,b}$}
\email{liuyx@lzu.edu.cn}

\affiliation{$^{a}$ Lanzhou Center for Theoretical Physics, Key Laboratory of Theoretical Physics of Gansu Province, Key Laboratory of Quantum Theory and Applications of MoE, Gansu Provincial Research Center for Basic Disciplines of Quantum Physics, Lanzhou University, Lanzhou 730000, China \\
	$^{b}$ Institute of Theoretical Physics \& Research Center of Gravitation, School of Physical Science and Technology, Lanzhou University, Lanzhou 730000, China \\
	$^{c}$ Center for Gravitational Physics and Quantum Information,
	Yukawa Institute for Theoretical Physics,
	Kyoto University, Kyoto 606-8502, Japan \\
	$^{d}$ Research Center for the Early Universe (RESCEU),
	Graduate School of Science, The University of Tokyo,
	Hongo 7-3-1, Bunkyo-ku, Tokyo 113-0033, Japan \\
	$^{e}$ Kavli Institute for the Physics and Mathematics of the Universe (WPI),
	The University of Tokyo Institutes for Advanced Study,
	The University of Tokyo, Kashiwa, Chiba 277-8583, Japan}

\begin{abstract}
\textbf{Abstract:} We analyze the propagation and polarization properties of high-frequency gravitational waves in Einstein–\AE ther theory on vorticity-free and slowly-varying backgrounds at both leading and next-to-leading orders within the geometric optics approximation. The linear perturbation analysis is performed in the background \AE ther-orthogonal frame, in which the axes of the gravitational wave sound cones remain perpendicular to these hypersurfaces, thereby simplifying the analysis. The leading-order results show that Einstein–\AE ther theory admits two tensor modes, two vector modes, and one scalar mode, consistent with the findings in the flat spacetime background. We further derive the dispersion relations and linear stability conditions for these modes in curved backgrounds. At next-to-leading order, we obtain the amplitude evolution equations, finding that the graviton number is conserved for the tensor modes but not for the vector and scalar modes. Next-to-leading-order effects also induce mixing among polarization modes. Our study demonstrates that, after imposing the GW170817 constraint on the propagation speed of gravitational waves, the vector modes mixed with the leading-order tensor modes cannot be used to distinguish between general relativity and Einstein–\AE ther theory. On the other hand, the mixing between scalar modes and the leading-order tensor modes leads to distinct predictions in the two theories, providing a promising avenue to test Einstein–\AE ther gravity through the detection of polarization mixing in gravitational waves.

\end{abstract}
	
\maketitle

\section{Introduction}
\label{sec: intro} 
The successful detection of gravitational waves marks the advent of the era of multi-messenger gravitational wave astronomy \cite{Abbott1,Abbott2}. In addition to traditional electromagnetic observations, gravitational waves can also provide valuable information about astrophysical sources and the Universe, offering new ways to constrain gravity theories and cosmological models. In this context, the detection of gravitational wave polarization modes has attracted particular attention. Unlike general relativity, which predicts only two tensor modes ($+$ and $\times$), modified gravity theories generally allow for extra polarization modes of gravitational waves \cite{f(R,f(R2,Horndeski,generalized Proca,Y.Dong1,Y.Dong2,Y.Dong3,Y.Dong4}. Due to the rather strong experimental constraints on violations of the equivalence principle, in modified gravity one usually assumes that all standard model matter fields universally couple to one metric, as invoked by Einstein's equivalence principle, so that test particles follow geodesics \cite{A. Papapetrou}. In this case, %simple 
modified gravity theories in four-dimensional spacetime can have up to six independent polarization modes of gravitational waves \cite{Eardley}: besides the two tensor modes predicted by general relativity, there can be up to two vector modes (vector-$x$ and vector-$y$) and up to two scalar modes (breathing and longitudinal). %More general modified gravity theories yet satisfying the Einstein's equivalence principle, including bigravity, multi-gravity and scalar-tensor gravity with multiple scalars, may propagate more modes in the gravity sector. 
Therefore, the detection of extra polarization modes provides a model-independent test of gravity beyond general relativity. Any observation of extra modes in current or future detectors \cite{J. Aasi,F. Acernese,T. Akutsu,M. Punturo,D. Reitze,P. Amaro-Seoane,Z. Luo,J. Luo,H. Xu,F. Jenet,B. W. Stappers,R. N. Manchester,B. C. Joshi} would offer unambiguous evidence for the need to modify general relativity.

However, the existence of additional polarization solutions in a theory does not guarantee that these modes are excited in realistic astrophysical scenarios \cite{Jose Maria Ezquiaga,James Healy}. Whether they are radiated depends on the way how the so-called screening mechanisms are at work in the vicinity of the source. Therefore, in the absence of a sufficiently complete understanding of the screening mechanisms for time-dependent astrophysical sources without high degrees of symmetry, the non-detection of extra polarization modes does not rule out modified gravity theories, and it is difficult to determine a priori whether and how a real gravitational wave event will radiate additional polarization modes.

A complementary approach is to study the propagation of gravitational waves on a given spacetime background. Typically, this is done by solving the linearized perturbation equations \cite{Isaacson1,Isaacson2}. For backgrounds with spatial homogeneity and isotropy, such as Minkowski or flat Friedmann-Robertson-Walker (FRW) spacetimes, a scalar–vector–tensor (SVT) decomposition usually enables the decoupling of the linearized equations \cite{James M. Bardeen,Weinberg}. This allows independent analysis of each decoupled sector \cite{Antonio De Felice1,Tsutomu Kobayashi,Antonio De Felice2}. However, on scales relevant for gravitational wave propagation within galaxies or for their detection in the Solar System, the background generally lacks such symmetry. This prevents complete decoupling and may cause polarization mixing \cite{Alice Garoffolo,Charles Dalang,Kei-ichiro Kubota}. The degree of mixing varies across modified gravity theories, providing a new way to test them observationally. Once sufficient measurement precision is achieved, this method could be highly effective. Even if a gravitational wave originates from a source emitting only the two tensor polarization modes, mixed vector and scalar modes—induced by background inhomogeneities between the source and the detector—might still be used to constrain the parameter space of candidate gravity theories.

Besides the motivation mentioned above, analyzing gravitational wave propagation on curved backgrounds is theoretically important. Gravitational waves carry not only source information but also information about the spacetime geometry through which they have propagated. Accurate waveform modeling must therefore account for propagation effects, which can also inform the distribution of matter and dark energy in the Universe, although effects from cold dark matter may be small \cite{Raphael Flauger}.

Motivated by these considerations, this work studies the propagation of gravitational waves on curved spacetime backgrounds in Einstein–\AE ther theory. Studies in quantum gravity suggest that Lorentz symmetry may be violated at high energies \cite{Giovanni Amelino-Camelia}. For example, a renormalizable, unitary and asymptotically free theory of quantum gravity called Ho\v{r}ava gravity~\cite{Horava:2009uw} breaks Lorentz symmetry at the fundamental level (see e.g. Ref. \cite{Mukohyama:2010xz} for review). To model this kind of situation, Einstein–\AE ther theory introduces a unit timelike \AE ther vector field $u^a$, defining a preferred time direction at each spacetime point \cite{Ted Jacobson1}. The unit constraint is imposed via a scalar Lagrange multiplier $\lambda$, preserving general covariance. Previous analyses of gravitational wave propagation in flat spacetime \cite{Ted Jacobson2} reveal that the theory supports five propagating degrees of freedom. These consist of two tensor and two vector modes, along with a single scalar degree of freedom. 
Due to Lorentz violation in this theory, gravitational waves generically propagate at speeds different from light. For a systematic analysis of the theoretical and observational constraints on Einstein–\AE ther theory, taking into account the stringent constraint on the propagation speed of tensor gravitational waves due to the gravitational wave event GW170817 and the gamma-ray burst GRB 170817A, see Ref. \cite{Jacob Oost}.

To study the propagation of these five degrees of freedom on general backgrounds, one first needs well-defined notions of tensor, vector, and scalar perturbations on such backgrounds. This can be achieved by extending the SVT decomposition to curved backgrounds \cite{Kei-ichiro Kubota}. The propagation properties can then be analyzed systematically, order by order, within the geometric optics approximation \cite{Isaacson1,Isaacson2,MTW,Michele Maggiore}. For simplicity, we consider cases where the background \AE ther vector field $\bar{u}^a$ is hypersurface-orthogonal, i.e., $\bar{u}^a$ has vanishing vorticity and there exists a scalar function $t(x^a)$ with $\bar{u}_a \propto \partial_a t$. This assumption holds for FRW spacetimes and static/dynamical spherically symmetric configurations \cite{Shinji Mukohyama1,Bhattacharjee:2018nus}. At the background level, this assumption also makes Einstein–\AE ther theory equivalent to the IR limit of the non-projectable version of Ho\v{r}ava gravity \cite{Ted Jacobson3}. In this case, it is convenient to adopt the \AE ther-orthogonal frame \cite{Shinji Mukohyama1} at the background level, where the background \AE ther vector is everywhere orthogonal to the $t=const$ hypersurfaces. In such a frame, the axes of the gravitational wave sound cones at each spacetime point are orthogonal to these hypersurfaces, which greatly simplifies the analysis and makes physical/geometrical interpretation manifest.

The paper is organized as follows. In Sec. \ref{sec: 2}, we briefly review Einstein–\AE ther theory. In Sec. \ref{sec: 3}, we introduce the SVT decomposition and the geometric optics approximation, which provide the essential tools for our analysis. In Sec. \ref{sec: 4}, we study the leading-order equations of gravitational waves within the geometric optics approximation. This allows us to identify the propagating degrees of freedom and their corresponding dispersion relations. We also examine the linear stability of perturbations on curved backgrounds, thereby generalizing results previously obtained for flat spacetime \cite{Jacob Oost}, cosmological background \cite{Noela Farina Sierra} and static spherically symmetric black hole backgrounds \cite{Shinji Mukohyama1,Antonio De Felice3}. Sec. \ref{sec: 5} investigates next-to-leading-order effects, deriving the amplitude evolution equations for each degree of freedom and identifying the mixed modes associated with the leading-order ones. In Sec. \ref{sec: 6}, we examine how these mixed modes are related to the gravitational wave polarization modes. Sec. \ref{sec: 7} presents the conclusion.

We use units with $c=1$ and adopt the metric signature $(-,+,+,+)$. The indices $(a,b,c,d,e,f)$ run over four-dimensional spacetime coordinates $(0,1,2,3)$. The indices $(i,j,k,l)$ run over three-dimensional spatial coordinates $(1,2,3)$, corresponding to the $(+x,+y,+z)$ directions, respectively. All background values of the corresponding quantities are denoted with an overbar, e.g., $\bar{u}^{a}$.

\section{Einstein-\AE ther theory}
\label{sec: 2}

In this section, we review Einstein–\AE ther theory. In this theory, gravity is described by the metric and an additional unit timelike vector field $u^{a}$, with the action given by \cite{Ted Jacobson2}
\begin{eqnarray}
	\label{action of Einstein-AEther theory}
	S=\frac{1}{16\pi G}\int d^{4}x \sqrt{-g}~
	\left[
	R
	+\mathcal{L}_{u}
	+\lambda\left(g_{ab}u^{a}u^{b}+1\right)
	\right].
\end{eqnarray}	
Here, $\lambda$ is a Lagrange multiplier,
\begin{eqnarray}
	\label{Lu}
	\mathcal{L}_{u}
	\coloneqq
	-M^{ab}_{~~cd}\nabla_{a}u^{c}\nabla_{b}u^{d},
\end{eqnarray}	
and
\begin{eqnarray}
	\label{Mabcd}
	M^{ab}_{~~cd}
	\coloneqq
	c_{1}g^{ab}g_{cd}
	+c_{2}\delta^{a}_{~c}\delta^{b}_{~d}
	+c_{3}\delta^{a}_{~d}\delta^{b}_{~c}
	-c_{4}u^{a}u^{b}g_{cd},
\end{eqnarray}	
where $c_{1}, c_{2}, c_{3}, c_{4}$ are dimensionless theoretical parameters, and $\delta^{a}_{~b}$ denotes the Kronecker delta.

Varying the action with respect to $\lambda$, we obtain
\begin{eqnarray}
	\label{u2=-1}
	g_{ab}u^{a}u^{b}=-1.
\end{eqnarray}	
This ensures that the vector field is timelike and normalized. Varying the action with respect to $u^{a}$, we have
\begin{eqnarray}
	\label{vector field equation}
	\mathcal{N}_{a}\left[g_{ab}, u^{a}\right]=
	\nabla_{b}J^{b}_{~a}+\lambda u_{a}+c_{4}(u^{b}\nabla_{b}u^{c})(\nabla_{a}u_{c})=0,
\end{eqnarray}	
where
\begin{eqnarray}
	\label{Jab}
	J^{a}_{~b}
	\coloneqq
	M^{ac}_{~~bd}\nabla_{c}u^{d}.
\end{eqnarray}	
Contracting Eq. (\ref{vector field equation}) with $u^{a}$, which gives $u^{a}\mathcal{N}_{a}=0$, and employing Eq. (\ref{u2=-1}), we can solve for $\lambda$ as
\begin{eqnarray}
	\label{Lambda=}
	\lambda=u^{a}\nabla_{b}J^{b}_{~a}+c_{4}u^{a}(\nabla_{a}u_{b})(u^{c}\nabla_{c}u^{b}).
\end{eqnarray}	
Finally, varying the action with respect to $g^{ab}$ yields
\begin{eqnarray}
	\label{tensor field equation}
	\mathcal{M}_{ab}\left[g_{ab}, u^{a}\right]\!&=&\!
	R_{ab}-\frac{1}{2}g_{ab}R-\nabla_{c}
	\left[u_{(a}{J^{c}}_{b)}+u^{c}J_{(ab)}-a_{(a}{J_{b)}}^{c}\right]
	\nonumber \\
	\!&-&\!
	c_{1}
	\left[
	(\nabla_{a}u^{c})(\nabla_{b}u_{c})
	-(\nabla^{c}u_{a})(\nabla_{c}u_{b})
	\right]
	\nonumber\\
	\!&-&\!
	c_{4}(u^{c}\nabla_{c}u_{a})(u^{d}\nabla_{d}u_{b})
	-\frac{1}{2}g_{ab}\mathcal{L}_{u}
	-\lambda u_{a}u_{b}
	=0.
\end{eqnarray}	
In the following analysis, we always use Eq. (\ref{Lambda=}) to eliminate $\lambda$ from Eqs. (\ref{vector field equation}) and (\ref{tensor field equation}).

By decomposing $g_{ab}$ and $u^{a}$ into into background and perturbation parts, 
\begin{eqnarray}
	\label{g=g+h,A=A+a of vector-tensor theory 0}
	g_{ab}=\bar{g}_{ab}+h_{ab},~ u^{a}=\bar{u}^{a}+w^{a},
\end{eqnarray}	
where $\left| h_{ab} \right| \sim \left| w^{a} \right| \sim h \ll 1$, we can obtain the linearized perturbation equations corresponding to Eqs. (\ref{u2=-1}), (\ref{vector field equation}), and (\ref{tensor field equation}). These equations provide the foundation for analyzing gravitational wave propagation and polarization on curved backgrounds.

\section{SVT decomposition and the geometric optics approximation}
\label{sec: 3}

In this section, we introduce the essential tools needed to analyze gravitational wave propagation and polarization on curved backgrounds within the geometric optics approximation. For further details on this method, see Appendix \ref{app: Theoretical framework for analyzing gravitational wave propagation on arbitrary backgrounds}, where we systematically and comprehensively demonstrate how this approximation is applied to gravitational waves in modified gravity theories.

\subsection{Scalar-Vector-Tensor decomposition on curved spacetime backgrounds}
\label{sec: 2.1}
When solving the linearized perturbation equations on Minkowski or flat FRW backgrounds, the standard approach is to use a SVT decomposition \cite{James M. Bardeen,Weinberg}, which separates the perturbations into a transverse traceless spatial tensor, transverse spatial vectors, and spatial scalars. This decomposition allows us to discuss the concept of tensor, vector, and scalar mode gravitational waves, thereby providing a direct and clear physical picture of their propagation and polarization \cite{Y.Dong2,Y.Dong3}. From this viewpoint, tensor, vector, and scalar mode gravitational waves are defined by the SVT decomposition. To extend this physical picture to curved spacetime backgrounds and continue describing gravitational waves in terms of tensor, vector, and scalar modes, the SVT decomposition must be generalized to curved backgrounds (see, e.g., \S~II-1 and Appendix~B of Ref. \cite{Hideo Kodama}).

To introduce the SVT decomposition on a curved spacetime background, we begin by decomposing the background spacetime, in the region of interest, into a family of non-overlapping three-dimensional spatial slices $\Sigma_{t}$, and introduce a background unit timelike vector field 
$\bar{n}^{a}$ (with $\bar{g}_{ab}\bar{n}^a\bar{n}^b=-1$) that is orthogonal to $\Sigma_{t}$ at every point. (Later we shall set $\bar{n}^a=\bar{u}^a$.) In this construction, the induced metric on the spatial hypersurfaces $\Sigma_{t}$ is denoted by 
\begin{eqnarray}
	\label{induced metric gamma}
	\bar{\gamma}_{ab}\coloneqq \bar{g}_{ab}+\bar{n}_{a}\bar{n}_{b},
\end{eqnarray}
where $\bar{n}_a=\bar{g}_{ab}\bar{n}^b$. It is straightforward to show that $\bar{\gamma}_{ab}$ satisfies $\bar{\gamma}^{a}_{~c}\bar{\gamma}^{c}_{~b}=\bar{\gamma}^{a}_{~b}$ and $\bar{\gamma}_{ab}\bar{n}^{b}=0$, and therefore acts as a projection operator. It can be used to project any tensor onto a spatial tensor defined on the hypersurface $\Sigma_{t}$. Thus, any vector $V^{a}$ can be decomposed into a temporal component along $\bar{n}^{a}$
and a purely spatial component orthogonal to $\bar{n}^{a}$:
\begin{eqnarray}
	\label{V=Vp+Vo}
	V^{a}=V_{\perp} \bar{n}^{a} + V_{\parallel}^{a},
\end{eqnarray}
where $V_{\perp}\coloneqq -\bar{n}_{a}V^{a}$, and $V_{\parallel}^{a} \coloneqq \bar{\gamma}^{a}_{~b}V^{b}$. For any rank-$2$ tensor $T_{ab}$, a similar decomposition holds,
\begin{eqnarray}
	\label{T=Tp+To}
	T_{ab}=\left(\bar{n}^{c}\bar{n}^{d}T_{cd}\right)\bar{n}_{a}\bar{n}_{b}
	+\left(-\bar{n}^{c}\bar{\gamma}_{b}^{~d}T_{cd}\right)\bar{n}_{a}
	+\left(-\bar{n}^{d}\bar{\gamma}_{a}^{~c}T_{cd}\right)\bar{n}_{b}
	+\left(\bar{\gamma}_{a}^{~c}\bar{\gamma}_{b}^{~d}T_{cd}\right).
\end{eqnarray}
By the same token, this decomposition applies to tensors of arbitrary rank. Using the spacetime orthogonal projection introduced above, the time–time, time–space, and space–space components of the metric perturbation $h_{ab}$ are denoted respectively by
\begin{eqnarray}
	\label{time–time, time–space, and space–space components of h}
	A \coloneqq \bar{n}^{a}\bar{n}^{b}h_{ab}, ~
	J_{a} \coloneqq -\bar{\gamma}_{a}^{~b}\bar{n}^{c}h_{bc}, ~
	F_{ab} \coloneqq \bar{\gamma}_{a}^{~c}\bar{\gamma}_{b}^{~d}h_{cd}.
\end{eqnarray}
Similarly, the time and space components of the vector field perturbation $w^{a}$ are denoted respectively by
\begin{eqnarray}
	\label{time–time, time–space, and space–space components of w}
	L \coloneqq -\bar{n}_{a}w^{a},~W^{a} \coloneqq \bar{\gamma}^{a}_{~b}w^{b}. 
\end{eqnarray}

Using the background induced metric $\bar{\gamma}_{ab}$, one can naturally define a spatial covariant derivative operator $\bar{D}_{a}$ from the covariant derivative $\bar{\nabla}_{a}$ that is compatible with the background metric $\bar{g}_{ab}$. The derivative operator $\bar{D}_{a}$ acts exclusively on spatial tensors defined on $\Sigma_{t}$ and maps them to other spatial tensors on $\Sigma_{t}$ at the same point. For any spatial tensor $S^{a...}_{~b...}$ on $\Sigma_{t}$, we have
\begin{eqnarray}
	\label{Def D}
	\bar{D}_{c}S^{a...}_{~b...} \coloneqq
	\left(\bar{\gamma}_{c}^{~f}\bar{\gamma}^{a}_{~c}\bar{\gamma}_{b}^{~d}...\right)
	\left(\bar{\nabla}_{f}S^{c...}_{~d...}\right).
\end{eqnarray}
It is easy to see that $\bar{D}_{a}$ is compatible with $\bar{\gamma}_{ab}$: $\bar{D}_{c}\bar{\gamma}_{ab}=0$.

With the above geometric structures in place, we can now naturally define the SVT decomposition on a curved spacetime background as follows \cite{Kei-ichiro Kubota}:
\begin{eqnarray}
	W^{a}\!&=&\!\bar{D}^{a}M+N^{a},
	\nonumber\\
	\label{SVT decomposition in arbitrary backgrounds}
	J_{a}\!&=&\!\bar{D}_{a}B+B_{a},
	\\
	\nonumber
	F_{ab}\!&=&\!
	E_{ab}+\bar{D}_{a}E_{b}+\bar{D}_{b}E_{a}+\bar{\gamma}_{ab}C+\left(\bar{D}_{a}\bar{D}_{b}-\frac{1}{3}\bar{\gamma}_{ab}\bar{\gamma}^{cd}\bar{D}_{c}\bar{D}_{d}\right)E.
\end{eqnarray}
Here,
\begin{eqnarray}
	\label{transverse and traceless conditation}
	&\bar{D}_{a}N^{a}=\bar{D}^{a}B_{a}=\bar{D}^{a}E_{a}=0,  \nonumber\\
	&\bar{\gamma}^{ab}E_{ab}=\bar{D}^{a}E_{ab}=0.
\end{eqnarray}
According to the theory of second-order elliptic operators \cite{Eberhard Zeidler,Evans}, imposing an appropriate boundary condition such as the Dirichlet boundary condition $u|_{\partial\Omega}=0$ on the boundary of the region under consideration $\Omega$ ensures that the equation $\bar{g}^{ab}\bar{D}_{a}\bar{D}_{b}u=f$ admits a unique solution, where $u$ is a scalar function and $f$ is any sufficiently regular source term. Therefore, the SVT decomposition of a vector field is guaranteed to exist and is unique. A similar argument can be extended to tensor fields, ensuring that their decomposition also exists and is unique. This ensures that the tensor, vector, and scalar modes of gravitational waves are well-defined on a curved background.

For Einstein–\AE ther theory considered in this paper, we restrict our analysis to scenarios that satisfy the following assumptions: $\bar{u}^a$ is hypersurface-orthogonal, i.e. $\bar{u}^a$ has vanishing vorticity and there exists a scalar function $t(x^a)$ such that $\bar{u}_a \propto \partial_a t$. Accordingly, we can define spatial slices $\Sigma_{t}$ that are everywhere orthogonal to $\bar{u}^{a}$ in the region of interest, and, throughout the main text of this paper, we consistently set $\bar{n}^{a}\equiv\bar{u}^{a}$. This type of spacetime foliation is known as the background \AE ther-orthogonal frame \cite{Shinji Mukohyama1, Antonio De Felice3}. In this case, $\bar{\gamma}_{ab}\bar{u}^{b}=0$. As we will see below, after applying the SVT decomposition to the linearized perturbation equations, this condition ensures decoupling at the leading order in the geometric optics approximation. Specifically, the equations describing the spatial tensor, vector, and scalar modes from the SVT decomposition (\ref{SVT decomposition in arbitrary backgrounds}) become independent of each other (we provide a detailed demonstration of this in Appendix \ref{sec: A1}). This greatly simplifies the equations. Moreover, the setup also ensures that at every spacetime point, the axis of the gravitational wave sound cone is orthogonal to  the spatial slice $\Sigma_{t}$. As a result, the dispersion relations of gravitational waves are straightforward.

Finally, we assume that $\bar{g}_{ab}$ and $\bar{u}^{a}$ satisfy the field equations. This implies that, after substituting Eq. (\ref{Lambda=}), the temporal component of the linearized perturbation equation corresponding to Eq. (\ref{vector field equation}) along $\bar{u}^{a}$  vanishes identically. Therefore, we do not use this equation in the subsequent analysis. One might worry that this reduces the number of independent equations. However, it should be noted that we still have the constraint equation (\ref{u2=-1}). At linear order in the perturbations, Eq. (\ref{u2=-1}) requires the $3+1$ decompositions (\ref{time–time, time–space, and space–space components of h})-(\ref{time–time, time–space, and space–space components of w}) to satisfy 
\begin{eqnarray}
	\label{A=2L}
	A=2L.
\end{eqnarray}

\subsection{Geometric optics approximation}
\label{sec: 2.2}
Even after carrying out the SVT decomposition, directly solving the linear perturbation equations generally remains challenging, because the spatial scalar, vector, and tensor components on a curved spacetime background typically remain coupled. Nevertheless, employing the geometric optics approximation renders the problem tractable.

For the geometric optics approximation to be valid, two conditions must be satisfied. The first is that the reduced wavelength of the gravitational wave, defined as $\bar{\lambda} \coloneqq \lambda/2\pi$, with $\lambda$ being the wavelength, should be much smaller than the characteristic variation scale of the background spacetime, $\mathfrak{L}$, i.e. $\bar{\lambda}\ll \mathfrak{L}$. This condition ensures the applicability of the Isaacson picture for defining gravitational waves. The second condition requires the reduced wavelength $\bar{\lambda}$ to be much smaller than the typical variation scale $l$ of the gravitational-wave wavelength, amplitude, and polarization matrix; that is, $\bar{\lambda}\ll l$. This allows gravitational wave propagation to be treated in a manner analogous to rays in geometric optics.

Accordingly, we can introduce a single small parameter $\epsilon$ ($\epsilon \ll 1$), defined as the larger of $\bar{\lambda}/\mathfrak{L}$ and $\bar{\lambda}/l$, and use it to carry out a geometric optics expansion on the SVT decomposed perturbations:
\begin{eqnarray}
	\label{geometric optics approximation for SVT perturbations}
	A \!&=&\!\sum_{I=0}^{\infty}\overset{(I)}{A}\epsilon^{I}
	e^{i\left(\frac{\theta}{\epsilon}\right)},
	\quad
	B=\epsilon\sum_{I=0}^{\infty}\overset{(I)}{B}\epsilon^{I}e^{i\left(\frac{\theta}{\epsilon}\right)},
	\quad
	C=\sum_{I=0}^{\infty}\overset{(I)}{C}\epsilon^{I}e^{i\left(\frac{\theta}{\epsilon}\right)},
	\quad
	E=\epsilon^{2}\sum_{I=0}^{\infty}\overset{(I)}{E}\epsilon^{I}e^{i\left(\frac{\theta}{\epsilon}\right)},
	\nonumber \\
	L \!&=&\!\sum_{I=0}^{\infty}\overset{(I)}{L}\epsilon^{I}e^{i\left(\frac{\theta}{\epsilon}\right)},
	\quad
	M=\epsilon\sum_{I=0}^{\infty}\overset{(I)}{M}\epsilon^{I}e^{i\left(\frac{\theta}{\epsilon}\right)},
	\quad
	B_{a}=\sum_{I=0}^{\infty}\overset{(I)}{B_{a}}\epsilon^{I}e^{i\left(\frac{\theta}{\epsilon}\right)},
	\\
	\nonumber 
	E_{a}\!&=&\!\epsilon\sum_{I=0}^{\infty}\overset{(I)}{E_{a}}\epsilon^{I}e^{i\left(\frac{\theta}{\epsilon}\right)},
	\quad
	N^{a}=\sum_{I=0}^{\infty}\overset{(I)}{N^{a}}\epsilon^{I}e^{i\left(\frac{\theta}{\epsilon}\right)},
	\quad
	E_{ab}=\sum_{I=0}^{\infty}\overset{(I)}{E_{ab}}\epsilon^{I}e^{i\left(\frac{\theta}{\epsilon}\right)},
\end{eqnarray}
where $\theta/\epsilon$ is a real number characterizing the phase. To be more precise, we introduce the expansion (\ref{geometric optics approximation for SVT perturbations}) for each of scalar, vector and tensor sectors with an independent phase and amplitudes, and then consider the sum of them. Consequently, the four-dimensional wave vector can be defined as $k_{a}\coloneqq \partial_{a}\theta/\epsilon$. In the chosen spatial slices, $k_{a}$ can be decomposed as
\begin{equation}
	\label{k decomposed}
	k_{a} = \omega\, \bar{n}_{a} + {}^{(3)}k_{a} .
\end{equation}
Here, $\omega \coloneqq  -\bar{n}^{a}k_{a}$ is the frequency, and ${}^{(3)}k_{a} = \bar{\gamma}_{a}^{~b} k_{b}$ is the spatial wave vector. The amplitudes, such as $\overset{(I)}{E_{ab}}$ and $\overset{(I)}{N^{a}}$, are generally complex. This enables the geometric optics approximation to also describe gravitational waves with elliptical polarization. Finally, we note that in Eq. (\ref{geometric optics approximation for SVT perturbations}), some spatial vector and scalar amplitudes are not expanded with the zeroth power of $\epsilon$ as the leading order.  This is done so that the contributions of these quantities match the geometric optics approximation of $h_{ab}$ and $w^{a}$, since they enter $h_{ab}$ and $w^{a}$ through derivative terms.

In the geometric optics approximation, one can substitute (\ref{geometric optics approximation for SVT perturbations}) into the linearized perturbation equations and analyze the equations order by order. However, when the analysis is extended to higher orders in the geometric optics expansion, the linearized perturbation equations alone may become insufficient. This is because the $\mathcal{O}(h^{2})$ contributions in the high-frequency field equations can become comparable to the higher-order terms in the geometric optics expansion of the linearized perturbation equations. To avoid such situations we assume that $h\ll\epsilon\ll1$ \cite{Kei-ichiro Kubota}, so that the leading and next-to-leading orders of the geometric optics approximation considered in this work can be consistently studied using only the linearized perturbation equations.

\subsection{Gauge condition}
\label{sec: 2.3}
Imposing gauge conditions can further simplify the equations under analysis. Here, we choose to impose the following gauge condition:
\begin{eqnarray}
	\label{gauge condition}
	B=E=0,\quad E_{a}=0.
\end{eqnarray}
Under an infinitesimal coordinate transformation $x^{a}\rightarrow x^{a}+\xi^{a}$, where $\xi^{a}$ is an arbitrary high-frequency function satisfying $|\bar{\nabla}_{a}\xi_{b}|\sim\mathcal{O}(h)$, $h_{ab}$ transforms according to the following relation (to $\mathcal{O}(h)$):
\begin{eqnarray}
	\label{infinitesimal coordinate transformation of h}
	h_{ab} \rightarrow h_{ab}-\bar{\nabla}_{a}\xi_{b}-\bar{\nabla}_{b}\xi_{a}.
\end{eqnarray}
Considering only the SVT decompositions of the perturbation and $\xi^{a}$, and formally expanding $\xi^{a}$ in powers of $\epsilon$, one can use the uniqueness of the SVT decomposition together with Eqs. (\ref{geometric optics approximation for SVT perturbations}) and (\ref{infinitesimal coordinate transformation of h}) to show, order by order, that the gauge condition (\ref{gauge condition}) can always be imposed. This condition uniquely fixes the coordinates and leaves no residual gauge freedom.

After imposing this gauge, the perturbations can always be expressed with
\begin{eqnarray}
	\label{gauge perturbations SVT}
	w^{a}\!&=&\! L\bar{n}^{a}+\bar{D}^{a}M+N^{a}, \nonumber \\
	h_{ab}\!&=&\! A\bar{n}_{a}\bar{n}_{b}+B_{a}\bar{n}_{b}+B_{b}\bar{n}_{a}+C \bar{\gamma}_{ab}+E_{ab}.
\end{eqnarray}

From now on throughout the main text, as we have already stated, we shall set $\bar{n}^a=\bar{u}^a$.

\section{Leading-order analysis in Einstein-\AE ther theory}
\label{sec: 4}
In this section, we use the method described in Sec. \ref{sec: 3} to examine the propagation and polarization of gravitational waves in Einstein-\AE ther theory at the leading order within the geometric optics approximation. We also present the corresponding stability conditions at this order. At leading order in the geometric optics approximation, the tensor, vector, and scalar modes are completely decoupled from each other.

\subsection{Propagation speed and polarization modes of gravitational waves}
\label{sec: 4.1}
By substituting Eqs. (\ref{gauge perturbations SVT}) and (\ref{geometric optics approximation for SVT perturbations}) into the linearized perturbation equations corresponding to Eqs. (\ref{u2=-1}), (\ref{vector field equation}), and (\ref{tensor field equation}), and then projecting them on the transverse-traceless part, we analyze the tensor perturbations. At the leading order of the geometric optics approximation and find that they satisfy the equation
\begin{eqnarray}
	\label{tensor mode equation in leading order}
	\left[\left(-1+c_{13}\right)\omega^{2}+{}^{(3)}k^{2}\right]\overset{(0)}{E}_{ab}=0.
\end{eqnarray}	
Here, ${}^{(3)}k^{2}\coloneqq\bar{\gamma}_{ab}{}^{(3)}k^{a}{}^{(3)}k^{b}=\bar{g}_{ab}{}^{(3)}k^{a}{}^{(3)}k^{b}$ and $c_{13}\coloneqq c_{1}+c_{3}$. Similarly, in the following, for indexes $I, J, K=1, 2, 3, 4$, we set $c_{IJ}\coloneqq c_I+c_J$ and $c_{IJK}\coloneqq c_I+c_J+c_K$. It is clear that Einstein-\AE ther theory admits tensor-mode gravitational wave solutions, which possess two propagating degrees of freedom. In the background \AE ther-orthogonal frame, their propagation speed $v_{T}$ is given by
\begin{eqnarray}
	\label{vT2}
	v_{T}^{2}=\frac{1}{1-c_{13}}.
\end{eqnarray}	

For the vector perturbations, the equations are
\begin{eqnarray}
	\label{vector mode equation in leading order 1}
	&2\left[c_{1}{}^{(3)}k^{2}-c_{14}\omega^{2}\right]\overset{(0)}{N}_{a}
	+\left[
	\left(-c_{1}+c_{3}\right){}^{(3)}k^{2}
	+2c_{14}\omega^{2}
	\right]\overset{(0)}{B}_{a}
	=0,
	 \\
	\label{vector mode equation in leading order 2}
	&\left[
	\left(-1+c_1-c_3\right){}^{(3)}k^{2}
	-2c_{14}\omega^{2}
	\right]\overset{(0)}{B}_{a}
	+\left[
	\left(-c_1+c_3\right){}^{(3)}k^{2}
	+2c_{14}\omega^{2}
	\right]\overset{(0)}{N}_{a}
	=0,
	 \\
	\label{vector mode equation in leading order 3}
	&\overset{(0)}{B}_{a}-c_{13}\overset{(0)}{N}_{a}=0.
\end{eqnarray}	
These equations are not independent. By using Eq. (\ref{vector mode equation in leading order 3}), Eqs. (\ref{vector mode equation in leading order 1}) and (\ref{vector mode equation in leading order 2}) can both be reduced to
\begin{eqnarray}
	\label{vector mode equation in leading order 4}
	\left[
	\left(2c_1-c_1^2+c_3^2\right){}^{(3)}k^{2}
	-{2c_{14}(1-c_{13})}\omega^2
	\right]\overset{(0)}{N}_{a}
	=0.
\end{eqnarray}	
Therefore, the vector modes also carry two propagating degrees of freedom. In the background \AE ther-orthogonal frame, their propagation speed $v_{V}$ satisfies
\begin{eqnarray}
	\label{vV2}
	v_{V}^{2}=\frac{2c_1-c_1^2+c_3^2}{2c_{14}(1-c_{13})}.
\end{eqnarray}	
Moreover, their amplitudes obey Eq. (\ref{vector mode equation in leading order 3}).

Finally, the scalar perturbations satisfy the following equations:
\begin{eqnarray}
	\label{scalar mode equation in leading order 1}
	&-\left(c_{13}+3c_2\right)\omega \overset{(0)}{C}
	-2c_{14}\omega \overset{(0)}{L}
	+2ic_{123}{}^{(3)}k^{2}\overset{(0)}{M}
	-2ic_{14}\omega^2 \overset{(0)}{M}
	=0,
	\\
	\label{scalar mode equation in leading order 2}
	&{}^{(3)}k^{2}
	\big(
	\overset{(0)}{C}-c_{14}\overset{(0)}{L}-ic_{14}\omega\overset{(0)}{M}
	\big)
	=0,
	\\
	\label{scalar mode equation in leading order 3}
	&\omega
	\big(
	\overset{(0)}{C}-c_{14}\overset{(0)}{L}-ic_{14}\omega\overset{(0)}{M}
	\big)
	=0,
	\\
	\label{scalar mode equation in leading order 4}
	&
	\left[
	-{}^{(3)}k^{2}
	+\left(2+c_{13}+3c_2\right)\omega^2
	\right]\overset{(0)}{C}
	+2{}^{(3)}k^{2}\overset{(0)}{L}
	-2ic_{2}{}^{(3)}k^{2}\omega\overset{(0)}{M}
	=0,
	\\
	\label{scalar mode equation in leading order 5}
	&
	\overset{(0)}{C}
	-2\overset{(0)}{L}
	-2ic_{13}\omega\overset{(0)}{M}
	=0.
\end{eqnarray}	
It is clear that Eqs. (\ref{scalar mode equation in leading order 2}) and (\ref{scalar mode equation in leading order 3}) are equivalent. Combining either one with Eq. (\ref{scalar mode equation in leading order 5}), we obtain that the scalar amplitudes satisfy 
\begin{eqnarray}
	\label{scalar mode equation in leading order 6}
    i\omega\overset{(0)}{M}
    \!&=&\!
   - \frac{2-c_{14}}{c_1+2c_3-c_4}\overset{(0)}{L},
    \\
    \label{scalar mode equation in leading order 7}
    \overset{(0)}{C}
    \!&=&\!
    \left[
    2-
    \frac{2c_{13}\left(2-c_{14}\right)}{c_1+2c_3-c_4}
    \right]\overset{(0)}{L}.
\end{eqnarray}	
By substituting Eqs. (\ref{scalar mode equation in leading order 6}) and (\ref{scalar mode equation in leading order 7}) into Eqs. (\ref{scalar mode equation in leading order 1}) and (\ref{scalar mode equation in leading order 4}), we find that both are equivalent to
\begin{eqnarray}
	\label{scalar mode equation in leading order 8}
	\left[
	\left(c_{13}-1\right)c_{14}
	\left(2+c_{13}+3c_{2}\right)\omega^{2}
	-c_{123}\left(c_{14}-2\right){}^{(3)}k^2
	\right]\overset{(0)}{L}
	=0.
\end{eqnarray}	
Therefore, the scalar perturbations have only one propagating degree of freedom. In the background \AE ther-orthogonal frame, their propagation speed $v_{S}$ satisfies
\begin{eqnarray}
	\label{vS2}
	v_{S}^{2}=\frac{c_{123}\left(c_{14}-2\right)}{\left(c_{13}-1\right)c_{14}\left(2+c_{13}+3c_{2}\right)}.
\end{eqnarray}	
Taken together, the propagation speeds of all modes in the background \AE ther-orthogonal frame on curved background is the same as that on the Minkowski background \cite{Ted Jacobson2}.

\subsection{Stability conditions}
\label{sec: 4.2}
In this subsection, we derive the gradient stability and no-ghost conditions for Einstein-\AE ther theory on curved spacetime backgrounds, within the leading-order geometric optics approximation.

To avoid gradient instabilities, the squared propagation speed of gravitational waves must remain positive. Consequently, from Eqs. (\ref{vT2}), (\ref{vV2}), and (\ref{vS2}), we obtain the conditions
\begin{eqnarray}
	\label{gradient stability conditions}
	c_{13} < 1,\quad 
	\frac{2c_1-c_1^2+c_3^2}{2c_{14}}>0,\quad
	\frac{c_{123}\left(2-c_{14}\right)}{c_{14}\left(2+c_{13}+3c_{2}\right)}>0.
\end{eqnarray}	

To establish the no-ghost conditions, the coefficients of the kinetic terms in the second-order perturbation action must be positive at the leading order of the geometric optics approximation. We perform our analysis in the background \AE ther-orthogonal frame. This coordinate system satisfies the criteria for a ``good'' coordinate system as discussed in Ref. \cite{Eugeny Babichev} in the context of scalar-tensor theories and in Ref. \cite{Shinji Mukohyama1} in the context of Einstein-\AE ther theory, and therefore allows for a correct determination of the no-ghost conditions.
We first substitute the solution (\ref{Lambda=}) for $\lambda$ into the action (\ref{action of Einstein-AEther theory}) and enforce constraint (\ref{u2=-1}), thereby obtaining an equivalent action $S'$. We then substitute the SVT-decomposed perturbations (\ref{gauge perturbations SVT}) (satisfying constraint (\ref{A=2L})) into $S'$ and expand the action to second order with respect to these perturbations. In the second-order perturbation action, all terms involving derivatives of the background can be neglected, as their variations do not affect the leading-order field equations in the geometric optics approximation. This further implies that, when performing integration by parts, one can disregard terms in which derivatives act on the background, keeping only those where derivatives act exclusively on the perturbations. For the same reason, all covariant derivatives can be replaced by ordinary derivatives. Furthermore, since the spatial tensor and vectors in the SVT decomposition satisfy the transverse conditions, the terms responsible for coupling between the spatial tensor, vectors, and scalars also do not contribute to the leading-order field equations. Consequently, the no-ghost conditions for the tensor, vector, and scalar modes can be examined independently.

For tensor modes, the corresponding second-order perturbation action is
\begin{eqnarray}
	\label{S2T}
	\overset{(2,T)}{S}=
	-\frac{1}{64\pi G}
	\int d^{4}x \sqrt{-\bar{g}}
	\left[
	\bar{g}^{ab}\bar{g}^{cd}\bar{g}^{ef}\partial_{a}E_{ce}\partial_{b}E_{df}
	+c_{13}\bar{u}^{a}\bar{u}^{b}\bar{g}^{cd}\bar{g}^{ef}\partial_{a}E_{ce}\partial_{b}E_{df}
	\right].
\end{eqnarray}	
It should be noted that, although the field equations derived from this action are not exactly the same as the linearized perturbation equations of Einstein-\AE ther theory, they are equivalent to Eq. (\ref{tensor mode equation in leading order}) at the leading order in the geometric optics approximation. Thus, the action can be employed to obtain the no-ghost condition of Einstein-\AE ther theory at this leading order.

In the \AE ther-orthogonal frame we have chosen, the background metric and the background vector field can always be expressed in the following coordinate form \cite{MTW}:
\begin{eqnarray}
	\label{aether-orthogonal frame 3+1}
	\bar{u}^{a}=\left(\frac{1}{N}, -\frac{N^{i}}{N}\right),\quad
	\bar{u}_{a}=\left(-N, \overset{\rightarrow}{0}\right),\quad
	\label{gab=}
	\begin{pmatrix}
		\bar{g}^{00} & \bar{g}^{0j}\\
		\bar{g}^{i0} & \bar{g}^{ij}
	\end{pmatrix}
	=
	\begin{pmatrix}
		-\frac{1}{N^2} & \frac{N^{j}}{N^2}\\
		\frac{N^{i}}{N^2} & \bar{g}^{ij}-\frac{N^{i}N^{j}}{N^2}
	\end{pmatrix}.
\end{eqnarray}	
Here, $N$ is the lapse function, and $N^{i}$ is the shift vector. At leading order in the geometric optics approximation, we can always introduce symmetric traceless tensors $e^{(+)}_{ab}$ and $e^{(\times)}_{ab}$, so that the solution for $E_{ab}$ can be written in the following form:
\begin{eqnarray}
	\label{geometric optics leading order of Eab}
	E_{ab}=
	\left(	
	E_{(+)}e^{(+)}_{ab}+E_{(\times)}e^{(\times)}_{ab}
	\right)
	e^{i\left(\frac{\theta}{\epsilon}\right)},
\end{eqnarray}
where
$e^{(+)}_{ab}e_{(+)}^{ab}=e^{(\times)}_{ab}e_{(\times)}^{ab}=1, e^{(+)}_{ab}e_{(\times)}^{ab}=0, \bar{u}^{a}e^{(+)}_{ab}=\bar{u}^{a}e^{(\times)}_{ab}={}^{(3)}k^{a}e^{(+)}_{ab}={}^{(3)}k^{a}e^{(\times)}_{ab}=0$. Thus, by substituting Eq. (\ref{aether-orthogonal frame 3+1}) and solution (\ref{geometric optics leading order of Eab}) into the leading-order quadratic  Hamiltonian associated with the action (\ref{S2T}), and applying the leading-order geometric optics approximation, we find that the kinetic term takes the form
\begin{eqnarray}
	\label{kinetic term of tensor}
	\frac{\sqrt{-\bar{g}}}{64\pi G N^2}\left(1-c_{13}\right)
	\sum_{\alpha=+, \times}\partial_{0}E_{(\alpha)}\partial_{0}E_{(\alpha)}.
\end{eqnarray}
Thus, the no-ghost condition is
\begin{eqnarray}
	\label{no-ghost condition of tensor}
	1-c_{13}>0.
\end{eqnarray}

The second-order perturbation action for the vector modes at the leading-order in the geometric optics approximation is
\begin{eqnarray}
	\label{S2V}
	\overset{(2,V)}{S}
	=
	\frac{1}{32\pi G}
	\int d^{4}x 
    \sqrt{-\bar{g}}
	\!&\times&\!
	\left[
	\left(1-c_{13}\right)\bar{g}^{ab}\bar{g}^{cd}\partial_{a}B_{c}\partial_{b}B_{d}
	\right.
	\nonumber \\
	\!&+&\!
	\left(1+c_{13}+2c_{4}\right)\bar{u}^{a}\bar{u}^{b}\bar{g}^{cd}\partial_{a}B_{c}\partial_{b}B_{d}
	\nonumber \\
	\!&-&\!
    2c_{1}\bar{g}^{ab}\bar{g}^{cd}\partial_{a}N_{c}\partial_{b}N_{d}
    +2c_{4}\bar{u}^{a}\bar{u}^{b}\bar{g}^{cd}\partial_{a}N_{c}\partial_{b}N_{d}	
    \nonumber \\
    \!&+&\!
    2\left(c_1-c_3\right)\bar{g}^{ab}\bar{g}^{cd}\partial_{a}B_{c}\partial_{b}N_{d}
    \nonumber \\
    \!&-&\!
    \left.
    2\left(c_{13}+2c_4\right)\bar{u}^{a}\bar{u}^{b}\bar{g}^{cd}\partial_{a}B_{c}\partial_{b}N_{d}
    \right].
\end{eqnarray}	
In the above action, the kinetic matrix is degenerate, implying the presence of non-dynamical variables. These give rise to constraints in the corresponding Hamiltonian, preventing a direct determination of the no-ghost conditions from the eigenvalues of the kinetic matrix. Conversely, to obtain the correct no-ghost condition, we redefine variables via linear combinations to identify the non-dynamical variables and their constraint equations, then substitute these back into the action to determine the no-ghost condition. At leading order in the geometric optics approximation, the vector modes satisfy Eqs. (\ref{vector mode equation in leading order 1})–(\ref{vector mode equation in leading order 3}), so the relevant constraint equations must take the form of Eq. (\ref{vector mode equation in leading order 3}) at this order. (Here, we use the term `equations' because each component of Eq. (\ref{vector mode equation in leading order 3}) is treated as an independent equation.) Substituting Eq. (\ref{vector mode equation in leading order 3}) back into action (\ref{S2V}) yields 
\begin{eqnarray}
	\label{S2VN}
	\overset{(2,V)}{S'}
	\!&=&\!
	-\frac{1}{32\pi G}
	\int d^{4}x 
	\sqrt{-\bar{g}}
	\left(1-c_{13}\right)
	\left[
	\left(2c_1-c_1^2+c_3^2\right)\bar{g}^{ab}\bar{g}^{cd}\partial_{a}N_{c}\partial_{b}N_{d}
	\right.
	\nonumber\\
    \!&+&\!
	\left.
	\left(c_1^2+c_3^2-2c_4+2c_3c_4+2c_1c_3+2c_1c_4\right)\bar{u}^{a}\bar{u}^{b}\bar{g}^{cd}\partial_{a}N_{c}\partial_{b}N_{d}	
	\right].
\end{eqnarray}	
By analogy with the tensor modes, the no-ghost condition for the vector modes is
\begin{eqnarray}
	\label{no-ghost condition of vector}
	c_{14}>0.
\end{eqnarray}

Finally, we consider the second-order perturbation action $\overset{(2,S)}{S}$ for the scalar mode. The situation is somewhat more involved for the scalar mode, because in the SVT decomposition (\ref{SVT decomposition in arbitrary backgrounds}), the relation between $w^{a}$ and $M$ involves a derivative. As a result, the terms in $\overset{(2,S)}{S}$ containing $M$ include derivative operators of order higher than two. Nevertheless, the number of time derivatives in each term of the action never exceeds two, consistent with the field equations of Einstein-\AE ther theory, where each term involves two derivative operators, i.e., $\overset{(2,S)}{S}\big[\dot{L},\bar{D}_{a}L, \dot{C},\bar{D}_{a}C, \bar{D}_{a}\dot{M},\bar{D}_{a}\bar{D}_{b}M\big]$. As the expression for $\overset{(2,S)}{S}$ is rather lengthy, we do not list it here. Inspecting the kinetic matrix formed by the coefficients of all terms in $\overset{(2,S)}{S}$ that contain two time derivative operators reveals that the matrix is both degenerate and diagonal. From this, we note that 
$L$ is not a dynamical variable.

The structure of the SVT decomposition implies that, at leading order in the geometric optics approximation, the equation obtained by varying the action $\overset{(2,S)}{S}$ with respect to $L$ coincides with Eq. (\ref{scalar mode equation in leading order 2}). This indicates that we may replace the variation-derived equation with the following one:
\begin{eqnarray}
	\label{varying S2S with respect to L}
	\Delta
	\big(
	C-c_{14}L+c_{14}\bar{u}^{a}\bar{\nabla}_{a}{M}
	\big)
    =0,
\end{eqnarray}	
as any difference between them arises only at the next-to-leading order, which is beyond the scope of our analysis. Here, $\Delta\coloneqq\bar{\gamma}^{ab}\bar{D}_{a}\bar{D_{b}}$. We assume that all perturbation variables vanish sufficiently rapidly at the boundaries under consideration. Under this assumption, Eq. (\ref{varying S2S with respect to L}) becomes equivalent to the following constraint: 
\begin{eqnarray}
	\label{L=}
	C-c_{14}L+c_{14}\bar{u}^{a}\bar{\nabla}_{a}{M}
	=0.
\end{eqnarray}	
Thus, Eq. (\ref{L=}) can be substituted back into $\overset{(2,S)}{S}$ to eliminate $L$, yielding the reduced action $\overset{(2,S)}{S'}\big[\dot{C},\bar{D}_{a}C, \bar{D}_{a}\dot{M},\bar{D}_{a}\bar{D}_{b}M\big]$. 

Similarly, inspecting the kinetic matrix in $\overset{(2,S)}{S'}$ reveals that it is also diagonal and degenerate, indicating that $M$ is not a dynamical variable. The structure of the SVT decomposition indicates that, at leading order in the geometric optics approximation, the equation obtained by varying the action $\overset{(2,S)}{S'}$ with respect to $M$ coincides with Eq. (\ref{scalar mode equation in leading order 1}), subject to the constraint imposed by Eq. (\ref{L=}). Thus, we obtain the constraint on $M$:
\begin{eqnarray}
	\label{M=}
    M=\frac{2+c_{13}+3c_2}{2c_{123}}\Delta^{-1}\left(\bar{u}^{a}\bar{\nabla}_{a} {C}\right).
\end{eqnarray}	
Noting that only $\bar{\nabla}_{a}$, $\bar{u}^{a}$, and $\bar{g}_{ab}$ carry indices in $\overset{(2,S)}{S'}$, the derivative operators can appear in only two forms: $\bar{u}^{a}\bar{\nabla}_{a}$ and $\Delta$. Furthermore, Eq. (\ref{aether-orthogonal frame 3+1}) implies that $\Delta$ does not involve any time derivative operators. Hence, the fact that $M$ is not a dynamical variable in $\overset{(2,S)}{S'}$ implies that, by integration by parts, each occurrence of $M$ can be associated with a $\Delta$ operator, allowing it to be eliminated using Eq. (\ref{M=}). Consequently, we obtain the final form of the action:
\begin{eqnarray}
	\label{S2S}
	\overset{(2,S)}{\tilde{S}}
	\!&=&\!
	\frac{1}{64\pi G}\frac{1}{c_{123}c_{14}}
	\int d^{4}x 
	\sqrt{-\bar{g}}
	\left[
	c_{123}\left(-2+c_{14}\right)\bar{g}^{ab}\partial_{a}C\partial_{b}C
	\right.
	\nonumber\\
	\!&-&\!
	\left.
	\big(2\left(c_{2}+c_{3}-c_{4}\right)+\left(-4c_{2}+c_{13}\left(c_{13}+3c_{2}\right)\right)c_{14}\big)\bar{u}^{a}\bar{u}^{b}\partial_{a}C\partial_{b}C
	\right].
\end{eqnarray}	
It is evident that the no-ghost condition for the scalar mode is
\begin{eqnarray}
	\label{no-ghost condition of scalar}
	\frac{\left(1-c_{13}\right)\left(2+c_{13}+3c_{2}\right)}{c_{123}}>0.
\end{eqnarray}

Combining the gradient stability and no-ghost conditions (\ref{gradient stability conditions}), (\ref{no-ghost condition of tensor}), (\ref{no-ghost condition of vector}), and (\ref{no-ghost condition of scalar}), we ultimately find that the theory parameters must satisfy
\begin{eqnarray}
	\label{parameters condition}
	c_{13} < 1,\quad 
	0< c_{14} <2, \quad
	2c_1-c_1^2+c_3^2>0,\quad
	c_{123}\left(2+c_{13}+3c_{2}\right)>0.
\end{eqnarray}
By performing a leading-order analysis within the geometric optics approximation, we have established the linear stability conditions for Einstein-\AE ther theory on curved backgrounds. It can be seen that the condition in Eq. (\ref{parameters condition}) coincides with that in the Minkowski background \cite{Jacob Oost}. This is unsurprising, since no derivatives of the background fields enter our analysis at this order. However, when applying a similar procedure to Horndeski or generalized Proca theories, a similar analysis is expected to provide more information than in the Minkowski background with constant additional fields.

Finally, we add a brief remark beyond the main scope of this discussion. When analyzing stability conditions, we essentially require that the Hamiltonian of the system, or equivalently its energy, is bounded from below. In addition to constructing the Hamiltonian from the second-order perturbation action, there is another way to define the energy of gravitational waves (or perturbations), namely through the effective energy–momentum tensor in the Isaacson picture. A natural question then arises. Given the definition of gravitational wave energy in this latter approach, would requiring it to be bounded from below yield the same stability conditions? Reference \cite{Alexander Saffer} provides the expression for the effective energy–momentum tensor of gravitational waves in Einstein-\AE ther theory on the Minkowski background. This expression shows that requiring the effective energy of gravitational waves to be bounded from below, together with the gradient stability condition (\ref{gradient stability conditions}), is equivalent to imposing the condition (\ref{parameters condition}).

\section{Next-to-leading-order analysis in Einstein-\AE ther theory}
\label{sec: 5}
In this section, we separately investigate the next-to-leading-order effects in the geometric optics approximation for the tensor, vector, and scalar modes. Because the equations under consideration are linear, we can, without loss of generality, restrict our analysis to cases in which the leading-order mode is purely tensor, vector, or scalar. The next-to-leading-order effects of an arbitrary gravitational wave can then always be written as a linear combination of these results.

We first consider the case in which only the tensor modes contribute at the leading order in the geometric optics approximation. In this case, among the zeroth-order amplitudes defined in Eq. (\ref{geometric optics approximation for SVT perturbations}), only $\overset{(0)}{E_{ab}}$ is nonzero.

At this stage, the evolution equation for the tensor-mode amplitude is determined by the tensor part of the next-to-leading-order equations:
\begin{eqnarray}
	\label{tensor part of the next-to-leading-order Tensor modes}
	&\frac{1}{2}i\bar{\nabla}_{a}k^{a}\overset{(0)}{E_{fg}}
	+ik^{a}\bar{\nabla}_{a}\overset{(0)}{E_{fg}}
	-ic_{13}\omega\bar{u}^{a}\bar{\nabla}_{a}\overset{(0)}{E_{fg}}
	-\frac{1}{2}ic_{13}\bar{\nabla}_{a}\left(\omega\bar{u}^{a}\right)\overset{(0)}{E_{fg}}
	\nonumber\\
	&+ic_{13}\omega\left(\bar{u}^{a}\bar{\nabla}_{a}\bar{u}^{b}\right)\bar{u}_{(f}\overset{(0)~~}{E_{g)b}}
	+ic_{13}\omega\Big(\bar{\nabla}^{a}\bar{u}_{(f}\overset{(0)~~}{E_{g)a}}-\bar{\nabla}_{(f}\bar{u}^{a}\overset{(0)~~}{E_{g)a}}\Big)
	-2i\left(k^{b}\bar{\nabla}_{b}\bar{u}^{a}\right)\bar{u}_{(f}\overset{(0)~~}{E_{g)a}}
	\nonumber \\
	&-\frac{2}{{}^{(3)}k^2}{}^{(3)}k_{(f}
	\Big[
	-i\left(k^{a}\bar{\nabla}_{a}k^{b}\right)\overset{(0)~~}{E_{g)b}}
    +ic_{13}\omega\left(\bar{u}^{a}\bar{\nabla}_{a}k^{b}\right)\overset{(0)~~}{E_{g)b}}
	+\frac{1}{2}ic_{13}\omega\left(k^{a}\bar{\nabla}_{b}\bar{u}_{a}\right)\overset{(0)~~}{E_{g)b}}
	\nonumber \\
	&-\frac{1}{2}i\omega\left(c_{13}-2\right)\left(k^{a}\bar{\nabla}_{a}\bar{u}^{b}\right)\overset{(0)~~}{E_{g)b}}
	-\frac{1}{2}ic_{13}\omega^{2}\left(\bar{u}^{a}\bar{\nabla}_{a}\bar{u}^{b}\right)\overset{(0)~~}{E_{g)b}}
	\Big]
	=0.
\end{eqnarray}	
$\overset{(0)}{E_{ab}}$ can be expressed as $\overset{(0)}{E_{ab}}=E e_{ab}$, where 
$e_{ab}e^{ab}=1,\bar{u}^{a}e_{ab}={}^{(3)}k^{a}e_{ab}=0$. Therefore, substituting this form into the above equation and projecting both sides onto $E e^{fg}$ yields
\begin{eqnarray}
	\label{E2 tensor part of the next-to-leading-order Tensor modes}
	\frac{1}{2}\bar{\nabla}_{a}
	\left(
	ik^{a}E^2-ic_{13}\omega\bar{u}^{a}E^2
	\right)
	=0.
\end{eqnarray}	
This demonstrates the conservation of the number of tensor mode gravitons. Substituting Eq. (\ref{E2 tensor part of the next-to-leading-order Tensor modes}) back into Eq. (\ref{tensor part of the next-to-leading-order Tensor modes}) yields
\begin{eqnarray}
	\label{e tensor part of the next-to-leading-order Tensor modes}
	&
	ik^{a}\bar{\nabla}_{a}e_{fg}
	-ic_{13}\omega\bar{u}^{a}\bar{\nabla}_{a}e_{fg}
	\nonumber\\
	&=-ic_{13}\omega\left(\bar{u}^{a}\bar{\nabla}_{a}\bar{u}^{b}\right)\bar{u}_{(f}e_{g)b}
	-ic_{13}\omega\Big(\bar{\nabla}^{a}\bar{u}_{(f}{e_{g)a}}-\bar{\nabla}_{(f}\bar{u}^{a}{e_{g)a}}\Big)
	+2i\left(k^{b}\bar{\nabla}_{b}\bar{u}^{a}\right)\bar{u}_{(f}e_{g)a}
	\nonumber \\
	&+\frac{2}{{}^{(3)}k^2}{}^{(3)}k_{(f}
	\Big[
	-i\left(k^{a}\bar{\nabla}_{a}k^{b}\right){e_{g)b}}
    +ic_{13}\omega\left(\bar{u}^{a}\bar{\nabla}_{a}k^{b}\right){e_{g)b}}
	+\frac{1}{2}ic_{13}\omega\left(k^{a}\bar{\nabla}_{b}\bar{u}_{a}\right){e_{g)b}}
	\nonumber \\
	&-\frac{1}{2}i\omega\left(c_{13}-2\right)\left(k^{a}\bar{\nabla}_{a}\bar{u}^{b}\right){e_{g)b}}
	-\frac{1}{2}ic_{13}\omega^{2}\left(\bar{u}^{a}\bar{\nabla}_{a}\bar{u}^{b}\right){e_{g)b}}
	\Big].
\end{eqnarray}	
This gives the evolution equation for the polarization matrix $e_{ab}$.

For the vector part of the next-to-leading-order equations in the geometric optics approximation, the corresponding equations yield
\begin{eqnarray}
	\label{vector part of the next-to-leading-order Tensor modes 1}
	&\frac{1}{2}\left(c_1-c_3\right)k^2\epsilon\overset{(1)}{B_{g}}
	-\frac{1}{2} \left(c_{13}+2c_{4}\right)\omega^{2}\epsilon\overset{(1)}{B_{g}}
	-c_{1}k^2\epsilon\overset{(1)}{N_{g}}
	+c_{4}\omega^{2}\epsilon\overset{(1)}{N_{g}}
	\nonumber \\
	&=-ic_1k_{a}\bar{\nabla}^{b}\bar{u}^{a}\overset{(0)}{E_{gb}}
	-ic_3k^{a}\bar{\nabla}_{b}\bar{u}^{a}\overset{(0)}{E_{ga}}
	+\frac{1}{2}i\left(c_{13}+2c_{4}\right)\omega\bar{u}^{a}\bar{\nabla}_{a}\bar{u}^{b}\overset{(0)}{E_{gb}},
	\\
	\label{vector part of the next-to-leading-order Tensor modes 2}
	&\frac{1}{2}\left(-1+c_1-c_3\right)k^{2}\epsilon\overset{(1)}{B_{g}}
	-\frac{1}{2}\left(1+c_{13}+2c_{4}\right)\omega^{2}\epsilon\overset{(1)}{B_{g}}
	+\frac{1}{2}\left(-c_1+c_3\right)k^{2}\epsilon\overset{(1)}{N_{g}}
	\nonumber \\
	&+\frac{1}{2}\left(c_{13}+2c_{4}\right)\omega^2\epsilon\overset{(1)}{N_{g}}
	=\frac{1}{2}i\left(-1-c_{13}-2c_{4}\right)\omega^2\bar{u}^{a}\bar{\nabla}_{a}\bar{u}^{b}\overset{(0)}{E_{gb}}
	\nonumber \\
	&+\frac{1}{2}i\left(-c_1+c_3\right)k_{a}\bar{\nabla}^{b}\bar{u}^{a}\overset{(0)}{E_{gb}}
	+\frac{1}{2}i\left(-2+c_1-c_3\right)k_{b}\bar{\nabla}^{b}\bar{u}^{a}\overset{(0)}{E_{ga}},
	\\
	\label{vector part of the next-to-leading-order Tensor modes 3}	
	&\frac{1}{2}\omega\epsilon\overset{(1)}{B_{g}}
	-\frac{1}{2}c_{13}\omega\epsilon\overset{(1)}{N_{g}}
	=
	-\frac{1}{2}i\bar{u}^{b}\bar{\nabla}_{b}\bar{u}^{a}\overset{(0)}{E_{ga}}
	+\frac{1}{{}^{(3)}k^2}
	\Big[
	ik^{a}{}^{(3)}k^{b}\bar{\nabla}_{a}\overset{(0)}{E_{gb}}
	\nonumber \\
	&
	-ic_{13}\omega\bar{u}^{a}{}^{(3)}k^{b}\bar{\nabla}_{a}\overset{(0)}{E_{gb}}
	+\frac{1}{2}ic_{13}\omega{}^{(3)}k^{b}\bar{\nabla}^{a}\bar{u}_{b}\overset{(0)}{E_{ga}}
	-\frac{1}{2}ic_{13}\omega{}^{(3)}k^{b}\bar{\nabla}_{b}\bar{u}^{a}\overset{(0)}{E_{ga}}
	\Big]
	=0.
\end{eqnarray}	
From this, one can deduce that the mixed vector modes associated with the leading-order tensor modes satisfy
\begin{eqnarray}
	\label{secondary vector modes excited by the leading-order tensor modes B}	
	&\epsilon\overset{(1)}{B_{g}}
	=
	c_{13}\epsilon\overset{(1)}{N_{g}}
	-i\frac{\bar{u}^{b}\bar{\nabla}_{b}\bar{u}^{a}}{\omega}\overset{(0)}{E_{ga}}
	\nonumber \\
	&+\frac{2}{\omega{}^{(3)}k^2}
	\Big[
	-ik^{a}\bar{\nabla}_{a}k^{b}\overset{(0)}{E_{gb}}
	+i\omega k^{a}\bar{\nabla}_{a}\bar{u}^{b}\overset{(0)}{E_{gb}}
	+ic_{13}\omega\bar{u}^{a}\bar{\nabla}_{a}k^{b}\overset{(0)}{E_{gb}}
	\nonumber \\
	&
	+\frac{1}{2}ic_{13}\omega k^{a}\bar{\nabla}^{b}\bar{u}_{a}\overset{(0)}{E_{gb}}
	-\frac{1}{2}ic_{13}\omega k^{a}\bar{\nabla}_{a}\bar{u}^{b}\overset{(0)}{E_{gb}}
	-\frac{1}{2}ic_{13}\omega^{2}\bar{u}^{a}\bar{\nabla}_{a}\bar{u}^{b}\overset{(0)}{E_{gb}}
	\Big],
	\\
	\label{secondary vector modes excited by the leading-order tensor modes N}	
	&\epsilon\overset{(1)}{N_{g}}
	=
	\frac{i}{\omega{}^{(3)}k^2}\frac{1}{\left(-2c_1+c_1^2-c_3^2\right){}^{(3)}k^2+2\left(1-c_{13}\right)c_{14}\omega^2}\times
	\nonumber \\
	&\Big[-\omega\Big(
	-4c_4\omega^2+2c_3c_4\omega^2-c_1^2(k^2-\omega^2)+c_3^2(k^2+\omega^2)
	\nonumber \\
	&+2c_1(k^2+\omega^2+(-2+c_{34})\omega^2)	\Big)k_{b}\bar{\nabla}^{b}\bar{u}^{a}\overset{(0)}{E_{ga}}
	\nonumber\\
	&+\omega\left(
	2c_1c_{34}\omega^2-c_1^2(k^2-\omega^2)-2c_1(k^2+\omega^2)+c_3(2c_4\omega^2+c_3(k^2+\omega^2))
	\right)k_{a}\bar{\nabla}^{b}\bar{u}^{a}\overset{(0)}{E_{gb}}
	\nonumber \\
	&-2\left(2c_{14}\omega^2+(c_3-c_1)(k^2+\omega^2)\right)k^{a}\bar{\nabla}^{b}k_{a}\overset{(0)}{E_{gb}}
	\nonumber \\
	&+\Big(c_1^2\omega^2(k^2-\omega^2)+c_1(-2c_{34}\omega^4-\omega^2(k^2+\omega^2)+(k^2+\omega^2)^2)
	\nonumber \\
	&-c_3(2c_4\omega^2+(-1+c_3)\omega^2(k^2+\omega^2)+(k^2+\omega^2)^2)\Big)\bar{u}^{a}\bar{\nabla}_{a}\bar{u}^{b}\overset{(0)}{E_{gb}}
	\nonumber \\
	&+2c_{13}\omega(2c_{14}\omega^2+(c_3-c_1)(k^2+\omega^2))\bar{u}^{a}\bar{\nabla}^{b}k_{a}\overset{(0)}{E_{gb}}
	\Big].
\end{eqnarray}	
From the above expressions, it is evident that every term on the right-hand side contains $i$. This implies the presence of a phase difference between the mixed vector modes and the leading-order tensor modes. This behavior is readily understood. In the next-to-leading-order equations, requiring the contributions from the leading-order and next-to-leading-order amplitudes to be of the same order implies that, in the terms involving the latter, the exponential factor is acted on by one additional derivative compared with the terms involving the former. This leads to an extra factor of $ik_{a}$. From this perspective, not only the mixed vector modes but also the mixed scalar modes possess a phase difference relative to the leading-order tensor modes. Moreover, it is evident that the mixed modes are frequency-dependent.

For the scalar part of the next-to-leading-order equations, we find that
\begin{eqnarray}
	\label{scalar part of the next-to-leading-order Tensor modes 1}
	&
	\frac{1}{2}\left(c_{13}+3c_{2}\right)\omega\epsilon\overset{(1)}{C}
	+c_{14}\omega\epsilon\overset{(1)}{L}
	-ic_{123}k^2\epsilon\overset{(1)}{M}
	-i\left(c_{23}-c_{4}\right)\omega^2\epsilon\overset{(1)}{M}
	=\frac{1}{2}ic_{13}\bar{\nabla}^{b}\bar{u}^{a}\overset{(0)}{E_{ab}},
	\\
	\label{scalar part of the next-to-leading-order Tensor modes 2}
	&\left(k^2+\omega^2\right)\epsilon\overset{(1)}{C}
	-c_{14}\left(k^2+\omega^2\right)\epsilon\overset{(1)}{L}
	-ic_{14}\omega\left(k^2+\omega^2\right)\epsilon\overset{(1)}{M}
    =-\frac{1}{2}i\left(1-c_{13}\right)\omega\bar{\nabla}^{a}\bar{u}^{b}\overset{(0)}{E_{ab}},
    \\
    \label{scalar part of the next-to-leading-order Tensor modes 3}
    &-\omega\epsilon\overset{(1)}{C}
    +c_{14}\omega\epsilon\overset{(1)}{L}
    +ic_{14}\omega^{2}\epsilon\overset{(1)}{M}
    =\frac{1}{2}i\bar{\nabla}^{b}\bar{u}^{a}\overset{(0)}{E_{ab}},
    \\
    \label{scalar part of the next-to-leading-order Tensor modes 4}
    &\frac{1}{2}\overset{(1)}{C}
    -\overset{(1)}{L}
    -ic_{13}\omega\overset{(1)}{M}=0,
    \\
    \label{scalar part of the next-to-leading-order Tensor modes 5}
    &-\frac{1}{2}k^2\epsilon\overset{(1)}{C}
    +\frac{1}{2}\left(1+c_{13}+3c_{2}\right)\omega^2\epsilon\overset{(1)}{C}
    +\left(k^2+\omega^2\right)\epsilon\overset{(1)}{L}
    +ic_2\omega \left(k^2+\omega^2\right)\epsilon\overset{(1)}{M}
    \nonumber \\
    &
    =-\frac{1}{2}i\left(1-c_{13}\right)\omega\bar{\nabla}^{b}\bar{u}^{a}\overset{(0)}{E_{ab}}.
\end{eqnarray}	
Consequently, the mixed scalar mode amplitudes associated with the leading-order tensor modes satisfy
\begin{eqnarray}
	\label{secondary scalar modes excited by the leading-order tensor modes M}	
    &i\omega\epsilon\overset{(1)}{M}
    =-\frac{2-c_{14}}{c_1+2c_3-c_4}\epsilon\overset{(1)}{L}
    -\frac{1}{2}\frac{1}{c_1+2c_3-c_4}\frac{\bar{\nabla}^{b}\bar{u}^{a}}{\omega}\overset{(0)}{E_{ab}},
    \\
    \label{secondary scalar modes excited by the leading-order tensor modes C}	
    &\epsilon\overset{(1)}{C}
    =\left[
    2
    -\frac{2c_{13}\left(2-c_{14}\right)}{c_1+2c_3-c_4}
    \right]\epsilon\overset{(1)}{L}
    -i\frac{c_{13}}{c_1+2c_3-c_4}\frac{\bar{\nabla}^{b}\bar{u}^{a}}{\omega}\overset{(0)}{E_{ab}},
    \\
    \label{secondary scalar modes excited by the leading-order tensor modes L}	
    &\epsilon\overset{(1)}{L}
    =-i\Big[
    c_{13}\left(c_{1}+2c_{3}-c_{4}\right)\omega^2\bar{\nabla}^{b}\bar{u}^{a}\overset{(0)}{E_{ab}}
    -c_{123}k^2\bar{\nabla}^{b}\bar{u}^{a}\overset{(0)}{E_{ab}}
    \nonumber \\
    &+c_{13}\left(c_{13}+3c_{2}\right)\omega^2\bar{\nabla}^{b}\bar{u}^{a}\overset{(0)}{E_{ab}}
    -\left(c_{23}-c_{4}\right)\omega^2\bar{\nabla}^{b}\bar{u}^{a}\overset{(0)}{E_{ab}}\Big]
    \Big/
    \nonumber \\
    &\Big[
    2c_{123}\left(-2+c_{14}\right)\omega k^2
    +2\left(c_{23}-c_{4}\right)\left(-2+c_{14}\right)\omega^3
    \nonumber \\
    &-2c_{14}\left(c_1+2c_3-c_4\right)\omega^3
    -2c_{14}\left(-1+c_{13}\right)\left(c_{13}+3c_{2}\right)\omega^3
    \Big].
\end{eqnarray}	

We now turn to the next-to-leading-order analysis of the leading-order vector and scalar modes within the geometric optics approximation. For the leading-order vector and scalar modes, the evolution equations for their amplitudes, along with the expressions for the associated mixed modes, are rather lengthy. Therefore, we present these results in Appendix {\ref{app: Next-to-leading-order equations of vector and scalar modes}}. Unlike the tensor gravitons, whose number is conserved, the vector and scalar gravitons do not obey such conservation. In addition, a leading-order vector mode can be associated with mixed tensor and scalar modes, while a leading-order scalar mode can similarly be associated with mixed tensor and vector modes.

\section{Polarization modes of gravitational waves}
\label{sec: 6}

The discussions in Sec. \ref{sec: 4} and Sec. \ref{sec: 5} allow us to solve for the perturbations $h_{ab}$ and $w^{a}$. However, since these perturbations are gauge-dependent, they are not directly observable. This motivates a further analysis of their relation to the observable characteristics of gravitational waves, in particular, their connection to the physically meaningful polarization modes.

The polarization modes of gravitational waves are defined by the relative motion modes of test particles \cite{Eardley}. In four-dimensional spacetime, one can always locally adopt the proper detector frame at the location of the test particles \cite{Michele Maggiore}, such that the relative motion between two test particles obeys the following geodesic deviation equation:
\begin{eqnarray}
	\label{equation of geodesic deviation}
	\frac{d^{2}\eta_{i}}{dt^{2}}=-{R}_{i0j0}\eta^{j}.
\end{eqnarray}
Here, $\eta_{i}$ {represents} the relative displacement of the two test particles, and $R_{i0j0}$ denotes the corresponding component of the Riemann curvature tensor. As the motion of $\eta_{i}$ is entirely governed by ${R}_{i0j0}$, the six independent components of its leading-order, high-frequency part, $\overset{(1)}{R}_{i0j0}$, naturally define the six independent polarization modes of gravitational waves \cite{Y.Dong2}.

For our study, it is more convenient to express the above definition in a form that does not depend on any specific coordinate system. Assuming that the gravitational wave detector has a four-velocity $v^a$, we introduce at each point along its worldline three mutually orthonormal spacelike unit vectors $\tilde{e}^{a}_{(i)}$ $(i=1,2,3)$, all orthogonal to $v^{a}$. Together with $v^{a}$, these vectors form a complete spacetime tetrad. Here, $\tilde{e}^{a}_{(3)}$ is chosen to align with the spatial wave vector as measured by the detector, i.e., $\tilde{e}^{a}_{(3)} \propto k^{a}+\left(k^bv_b\right)v^{a}$. In this way, the polarization modes of gravitational waves can be defined as
\begin{eqnarray}
	\label{P1-P6}
	\mathcal{E}_{(i)(j)}\coloneqq
	\tilde{e}^{a}_{(i)}v^{b}\tilde{e}^{c}_{(j)}v^{d}\mathop{R}^{(1)}\!_{abcd}=\begin{pmatrix}
		P_{4}+P_{6} & P_{5} & P_{2}\\
		P_{5}       & -P_{4}+P_{6}  & P_{3}\\
		P_{2}       &  P_{3}   &   P_{1}
	\end{pmatrix}.
\end{eqnarray}
Here, $P_{4}$ and $P_{5}$ represent the familiar $+$ and $\times$ modes. $P_{2}$ and $P_{3}$ represent the vector-$x$ and vector-$y$ modes. Finally, $P_{1}$ and $P_{6}$ correspond to the longitudinal and breathing modes, respectively.

In Einstein–\AE ther theory, the four-velocity $v^{a}$ of a gravitational wave detector typically differs from the background \AE ther field $\bar{u}^{a}$. In this work, we assume that $v^{a}-\bar{u}^{a}$ is a small quantity, so that in the proper detector frame the spatial anisotropy induced by Lorentz violation in the vicinity of the detector remains weak. In the \AE ther-orthogonal frame, one can always choose, at each point in the spacetime region of interest, three unit spacelike vectors ${e}^{a}_{(i)}$ $(i=1,2,3)$ on $\Sigma_t$. Here, ${e}^{a}_{(3)} \propto {}^{(3)}k^{a}$. Together with $\bar{u}^{a}$, they form another spacetime tetrad. Since $v^a-\bar{u}^a$ is a small quantity, one can always choose appropriate directions for ${e}^{a}_{(1)}$ and ${e}^{a}_{(2)}$ so that the tetrads $\{v^a, \tilde{e}_{(i)}^a\}$ and $\{\bar{u}^a, {e}_{(i)}^a\}$ differ only by an infinitesimal Lorentz transformation:
\begin{eqnarray}
	\label{infinitesimal Lorentz transformation}
	v^{a}\!&=&\!\bar{u}^{a}+\alpha_1 e_{(1)}^a+\alpha_2 e_{(2)}^a+\alpha_3 e_{(3)}^a,
	\nonumber\\
	\tilde{e}_{(1)}^{a}\!&=&\! e_{(1)}^a+\alpha_1 \bar{u}^a+\alpha_4 e_{(2)}^a+\alpha_5 e_{(3)}^a,
	\nonumber \\
	\tilde{e}_{(2)}^{a}\!&=&\! e_{(2)}^a+\alpha_2 \bar{u}^a-\alpha_4 e_{(1)}^a+\alpha_6 e_{(3)}^a,
	\\
	\tilde{e}_{(3)}^{a}\!&=&\! e_{(3)}^a+\alpha_3 \bar{u}^a-\alpha_5 e_{(1)}^a-\alpha_6 e_{(2)}^a.
	\nonumber
\end{eqnarray}
Here, we use $\zeta$ to characterize the magnitude of $\alpha_1, ..., \alpha_6$, which satisfy $\zeta\ll 1$. Since $\tilde{e}^{a}_{(3)} \propto k^{a}+\left(k^bv_b\right)v^{a}$ and ${e}^{a}_{(3)} \propto {}^{(3)}k^{a}$, these six parameters are not independent. They satisfy the relation 
\begin{eqnarray}
	\label{relation alpha 1 to 6}
	\alpha_5=v_g \, \alpha_1,\quad
	\alpha_6=v_g \, \alpha_2,
\end{eqnarray}
where $v_g \coloneqq \omega/{}^{(3)}k$. Therefore, at leading order in the geometric optics approximation, when the gravitational wave is a tensor, vector, or scalar mode, the corresponding $v_g$ is given by $v_T$, $v_V$, and $v_S$ respectively. Finally, because the definitions of $e_{(1)}^a$, $e_{(2)}^a$, $\tilde{e}_{(1)}^a$, and $\tilde{e}_{(2)}^a$ allow some freedom, without loss of generality, we can choose $\alpha_2=\alpha_4=\alpha_6=0$. We therefore obtain
\begin{eqnarray}
	\label{infinitesimal Lorentz transformation a2=a4=a6=0}
	v^{a}\!&=&\!\bar{u}^{a}+\alpha e_{(1)}^a+\beta e_{(3)}^a,
	\nonumber\\
	\tilde{e}_{(1)}^{a}\!&=&\! e_{(1)}^a+\alpha \bar{u}^a+v_g \alpha e_{(3)}^a,
	\nonumber \\
	\tilde{e}_{(2)}^{a}\!&=&\! e_{(2)}^a,
	\\
	\tilde{e}_{(3)}^{a}\!&=&\! e_{(3)}^a+\beta \bar{u}^a-v_g \alpha e_{(1)}^a.
	\nonumber
\end{eqnarray}
Here, we relabel $\alpha_1$ and $\alpha_3$ as $\alpha$ and $\beta$, respectively. In what follows, for convenience, we formally assume that the contributions to the polarization modes induced by the deviation of $v^{a}$ from $\bar{u}^{a}$ are of the same order as the next-to-leading–order geometric optics corrections, i.e., $\zeta \sim \epsilon$. If this is not the case, one only needs to drop the subdominant terms in the relevant expressions.

By expanding the Riemann curvature tensor to linear order in the perturbation, one finds that
\begin{eqnarray}
	\label{R1i0j0}
	\mathop{R}^{(1)}\!_{abcd}
	=\frac{1}{2}\left(
	\bar{\nabla}_{c}\bar{\nabla}_{b}h_{ad}
	+\bar{\nabla}_{d}\bar{\nabla}_{a}h_{bc}
	-\bar{\nabla}_{c}\bar{\nabla}_{a}h_{bd}
	-\bar{\nabla}_{d}\bar{\nabla}_{b}h_{ac}
	+...
	\right),
\end{eqnarray}
where $...$ denotes terms that do not contribute to either the leading- or next-to-leading-order in the geometric optics approximation. From Eqs. (\ref{geometric optics approximation for SVT perturbations}), (\ref{gauge perturbations SVT}), and (\ref{infinitesimal Lorentz transformation a2=a4=a6=0}), it follows that the leading-order contribution of $\mathcal{E}_{(i)(j)}$ in the geometric optics approximation, $\overset{(1)}{\mathcal{E}}_{(i)(j)}$, satisfies
\begin{eqnarray}
	\label{leading order Eij}
	2 \overset{(1)}{\mathcal{E}}_{(i)(j)} e^{-i\frac{\theta}{\epsilon}}
	\!&=&\!
	\omega^{2}\,\overset{(0)}{E}_{(i)(j)}
	-\omega\, {}^{(3)}k_{(i)}\overset{(0)}{B}_{(j)}
	-\omega\, {}^{(3)}k_{(j)}\overset{(0)}{B}_{(i)}
	\nonumber \\
	\!&+&\!
	{}^{(3)}k_{(i)}{}^{(3)}k_{(j)}\overset{(0)}{A}
	+\omega^{2}\,\delta_{(i)(j)}\overset{(0)}{C}.
\end{eqnarray}
Here, the indices $(i), (j)$ are defined by contracting the corresponding tensor indices with the spatial vectors $e^{a}_{(i)}, e^{a}_{(j)}$, respectively, and $\delta_{(i)(j)}$ denotes the Kronecker delta. Since the leading-order amplitude satisfies the transverse and traceless conditions (\ref{transverse and traceless conditation of leading order}), Eq. (\ref{leading order Eij}) shows that the relation between the gravitational wave polarization modes and the leading-order amplitude is essentially the same as in the Minkowski background. The $+$ and $\times$ modes remain excited solely by the spatial tensor $\overset{(0)}{E_{ab}}$, the vector-$x$ and vector-$y$ modes remain excited only by the spatial vector $\overset{(0)}{B_{a}}$, and the breathing and longitudinal modes remain excited exclusively by the spatial scalars $\overset{(0)}{A}$ and $\overset{(0)}{C}$.

For the next-to-leading-order contribution $\overset{(2)}{\mathcal{E}}_{(i)(j)}$, we have
\begin{eqnarray}
	\label{next leading order Eij}
	2\overset{(2)}{\mathcal{E}}_{(i)(j)} e^{-i\frac{\theta}{\epsilon}}
	\!&=&\!
	2e^{-i\frac{\theta}{\epsilon}}\delta\tilde{e}^{a}_{(i)} \, \bar{u}^{b} \,  {e}^{c}_{(j)} \,  \bar{u}^{d}\mathop{R}^{(1,1)}\!\!_{abcd}
	+
	2e^{-i\frac{\theta}{\epsilon}} {e}^{a}_{(i)} \, \delta v^{b} \,  {e}^{c}_{(j)} \,  \bar{u}^{d}\mathop{R}^{(1,1)}\!\!_{abcd}
	\nonumber \\
	\!&+&\!
	2e^{-i\frac{\theta}{\epsilon}} {e}^{a}_{(i)} \,  \bar{u}^{b} \, \delta \tilde{e}^{c}_{(j)} \,  \bar{u}^{d}\mathop{R}^{(1,1)}\!\!_{abcd}
	+
	2e^{-i\frac{\theta}{\epsilon}} {e}^{a}_{(i)} \, \bar{u}^{b} \, {e}^{c}_{(j)} \, \delta v^{d}\mathop{R}^{(1,1)}\!\!_{abcd}
	\nonumber \\
	\!&+&\!
	\epsilon\, \omega^{2}\,\overset{(1)}{E}_{(i)(j)}
	-\epsilon\, \omega\, {}^{(3)}k_{(i)}\overset{(1)}{B}_{(j)}
	-\epsilon\, \omega\, {}^{(3)}k_{(j)}\overset{(1)}{B}_{(i)}
	\nonumber \\
	\!&+&\!
	\epsilon\,{}^{(3)}k_{(i)}{}^{(3)}k_{(j)}\overset{(1)}{A}
	+\epsilon\,\omega^{2}\,\delta_{(i)(j)}\overset{(1)}{C}
	\nonumber\\
	\!&-&\!
	i \omega \bar{u}^{c} e^{a}_{(i)}e^{b}_{(j)}\bar{\nabla}_{a}\overset{(0)}{E}_{bc}
	-i \omega \bar{u}^{c} e^{a}_{(i)}e^{b}_{(j)}\bar{\nabla}_{b}\overset{(0)}{E}_{ac}
	+2i \omega \bar{u}^{c} e^{a}_{(i)}e^{b}_{(j)}\bar{\nabla}_{c}\overset{(0)}{E}_{ab}
	\nonumber\\
	\!&+&\!
	i {}^{(3)}k_{(j)}
	\bar{u}^{b}\bar{u}^{c}e^{a}_{(i)}
	\bar{\nabla}_{c}\overset{(0)}{E}_{ab}
	+i {}^{(3)}k_{(i)}
	\bar{u}^{b}\bar{u}^{c}e^{a}_{(j)}
	\bar{\nabla}_{c}\overset{(0)}{E}_{ab}
	-i \bar{u}^{a}\bar{u}^{b}\bar{\nabla}_{b}k_{a}
	\overset{(0)}{E}_{(i)(j)}
	\nonumber \\
	\!&+&\!
	2i {}^{(3)}k_{(i)}\bar{u}^{b}e^{a}_{(j)}\bar{\nabla}_{a}\overset{(0)}{B}_{b}
	+2i {}^{(3)}k_{(j)}\bar{u}^{b}e^{a}_{(i)}\bar{\nabla}_{a}\overset{(0)}{B}_{b}
	\nonumber \\
	\!&-&\!
	i{}^{(3)}k_{(j)}\bar{u}^{b}e^{a}_{(i)}\bar{\nabla}_{b}\overset{(0)}{B}_{a}
	-i{}^{(3)}k_{(i)}\bar{u}^{b}e^{a}_{(j)}\bar{\nabla}_{b}\overset{(0)}{B}_{a}
	\nonumber \\
	\!&+&\!
	i\omega e^{a}_{(i)}e^{b}_{(j)}\bar{\nabla}_{a}\overset{(0)}{B}_{b}
	+i \omega
	e^{a}_{(i)}e^{b}_{(j)}\bar{\nabla}_{b}\overset{(0)}{B}_{a}
	\nonumber \\
	\!&-&\!
	i e^{a}_{(i)}\bar{u}^{b}\bar{\nabla}_{b}k_{a}
	\overset{(0)}{B}_{(j)}
	-i e^{a}_{(j)}\bar{u}^{b}\bar{\nabla}_{b}k_{a}
	\overset{(0)}{B}_{(i)}
	\nonumber \\
	\!&+&\!
	2i \omega
	e^{a}_{(i)}\bar{u}^{b}\bar{\nabla}_{b}\bar{u}_{a}
	\overset{(0)}{B}_{(j)}
	+2i \omega
	e^{a}_{(j)}\bar{u}^{b}\bar{\nabla}_{b}\bar{u}_{a}
	\overset{(0)}{B}_{(i)}
	\nonumber \\
	\!&-&\!
	i{}^{(3)}k_{(i)}e^{a}_{(j)}\bar{\nabla}_{a}\overset{(0)}{A}
	-i{}^{(3)}k_{(j)}e^{a}_{(i)}\bar{\nabla}_{a}\overset{(0)}{A}
	-i e^{a}_{(i)}e^{b}_{(j)}\bar{\nabla}_{b}k_{a}\overset{(0)}{A}
	\nonumber \\
	\!&-&\!
	i {}^{(3)}k_{(j)}e^{a}_{(i)}\bar{u}^{b}\bar{\nabla}_{b}\bar{u}_{a}\overset{(0)}{A}
	-i {}^{(3)}k_{(i)}e^{a}_{(j)}\bar{u}^{b}\bar{\nabla}_{b}\bar{u}_{a}\overset{(0)}{A}
	\nonumber \\
	\!&+&\!
	i\omega
	e^{a}_{(i)}e^{b}_{(j)}\bar{\nabla}_{a}\bar{u}_{b}\overset{(0)}{A}
	+i\omega
	e^{a}_{(i)}e^{b}_{(j)}\bar{\nabla}_{b}\bar{u}_{a}\overset{(0)}{A}
	\nonumber \\
	\!&-&\!
	i {}^{(3)}k_{(j)}e^{a}_{(i)}\bar{u}^{b}\bar{\nabla}_{b}\bar{u}_{a}\overset{(0)}{C}
	-i {}^{(3)}k_{(i)}e^{a}_{(j)}\bar{u}^{b}\bar{\nabla}_{b}\bar{u}_{a}\overset{(0)}{C}
	+2i\omega
	\delta_{(i)(j)}\bar{u}^{a}\bar{\nabla}_{a}\overset{(0)}{C}
	\nonumber \\
	\!&+&\!
	i \omega
	e^{a}_{(i)}e^{b}_{(j)}\bar{\nabla}_{a}\bar{u}_{b}\overset{(0)}{C}
	+i \omega
	e^{a}_{(i)}e^{b}_{(j)}\bar{\nabla}_{b}\bar{u}_{a}\overset{(0)}{C}
	-i \delta_{(i)(j)}\bar{u}^{a}\bar{u}^{b}\bar{\nabla}_{a}k_{b}\overset{(0)}{C}.
\end{eqnarray}
Here, $\delta v^{a} \coloneqq v^a-\bar{u}^a$, $\delta \tilde{e}^{a}_{(i)} \coloneqq \tilde{e}^{a}_{(i)}-{e}^{a}_{(i)}$, and $\overset{(1,1)}{R}\!\!_{abcd}$ denotes the leading-order term in the geometric optics expansion of $\overset{(1)}{R}_{abcd}$. Evidently, at next-to-leading order in the geometric optics expansion, the relation between the gravitational wave polarization modes and the tensor, vector, and scalar mode perturbations becomes more intricate. First, from the first four terms on the right-hand side, one can see that $v^a \neq \bar{u}^a$ leads to mixing among the different polarization modes. In addition, since the first-order amplitude in the geometric optics expansion (\ref{geometric optics approximation for SVT perturbations}) generally fails to satisfy the transverse condition (see Eq. (\ref{decomposition next-leading order break transverse conditation})), $\overset{(1)}{E}_{ab}$ no longer excites solely the $+$ and $\times$ modes, nor does  $\overset{(1)}{B}_{a}$ excite only the vector-$x$ and vector-$y$ modes. Finally, the appearance of zeroth-order amplitude terms in Eq. (\ref{next leading order Eij}) further complicates the correspondence between the perturbations and the polarization modes.

Since, to date, only the $+$ and $\times$ polarizations have been detected, we will therefore focus on the relationship between the mixed modes associated with the leading-order tensor modes and the gravitational wave polarization modes. Observations of the GW170817 event imply 
\cite{B.P.Abbott000,B.P.Abbott111}
\begin{eqnarray} 	
	\label{vT sim 1}
	-3\times10^{-15} \leq v_T-1 \leq 7\times10^{-16}.
\end{eqnarray}
Accordingly, we focus on the special case $v_T=1$, which corresponds to $c_{13}=0$. At this stage, among the next-to-leading order amplitudes in the geometric optics expansion, only $\overset{(1)}{E_{ab}}$, $\overset{(1)}{B_{a}}$, $\overset{(1)}{C}$ and $\overset{(1)}{L}$ contribute to Eq. (\ref{next leading order Eij}). From Eqs. (\ref{secondary vector modes excited by the leading-order tensor modes B}) (\ref{secondary scalar modes excited by the leading-order tensor modes C}) and (\ref{secondary scalar modes excited by the leading-order tensor modes L}), we obtain
\begin{eqnarray}
	\label{secondary vector modes excited by the leading-order tensor modes B c13=0}	
	&\epsilon\overset{(1)}{B_{g}}
	=
	-i\frac{1}{\omega}\bar{u}^{b}\bar{\nabla}_{b}\bar{u}^{a}\overset{(0)}{E_{ga}}
	+i \frac{2}{\omega^2}
	 k^{b}\bar{\nabla}_{b}\bar{u}^{a}\overset{(0)}{E_{ga}},
	 \\
	\label{secondary vector modes excited by the leading-order tensor modes C c13=0}	
	    &\overset{(1)}{C}
	=2\overset{(1)}{L},
	\\
	\label{secondary scalar modes excited by the leading-order tensor modes L c13=0}	
	&\epsilon\overset{(1)}{L}
	=i
	\frac{c_{2}-c_{14}}{4\left(c_{14}+2c_1c_2+c_2\left(-1+2c_4\right)\right)\omega}\bar{\nabla}^{b}\bar{u}^{a}\overset{(0)}{E_{ab}}.
\end{eqnarray}	
Here, we have used the relations $k^2=0$ and $k^{b}\bar{\nabla}_{b}k^a=0$. 

Substituting Eqs. (\ref{secondary vector modes excited by the leading-order tensor modes B c13=0})-(\ref{secondary scalar modes excited by the leading-order tensor modes L c13=0}) into Eq. (\ref{next leading order Eij}), we find that the first four terms on the right-hand side of Eq. (\ref{next leading order Eij}) vanish. Consequently, in the case considered here, the deviation of $v^a$ from $\bar{u}^a$ does not generate additional contributions to the gravitational wave polarization modes. This can also be understood from the fact that the $+$ and $\times$ modes propagating at the speed of light do not generate the other four polarization components under Lorentz transformations \cite{Eardley}. The remaining terms on the right-hand side of Eq. (\ref{next leading order Eij}) show that the mixed vector-$x$ mode associated with the leading-order tensor mode satisfies
\begin{eqnarray}
	\label{P2 for tensor mode}
	2\, P_2\, e^{-i\frac{\theta}{\epsilon}}
	\!&=&\!
	i \omega e_{(1)}^a \bar{\nabla}_{b}\overset{(0)}{E_{a}^{~b}}
	-i \omega e_{(1)}^a e_{(3)}^b \bar{\nabla}_{b}\bar{u}^c\overset{(0)}{E_{ac}}
	\nonumber \\
	\!&-&\!
	3i \omega e_{(1)}^a \bar{u}^b\bar{\nabla}_{b}\bar{u}^c\overset{(0)}{E_{ac}}
	+2i \omega e_{(1)}^a e_{(3)}^b \bar{u}^c \bar{\nabla}_c \overset{(0)}{E_{ab}}.
\end{eqnarray}
Similarly, for the vector-$y$ mode, we obtain
\begin{eqnarray}
	\label{P3 for tensor mode}
	2\, P_3\, e^{-i\frac{\theta}{\epsilon}}
	\!&=&\!
	i \omega e_{(2)}^a \bar{\nabla}_{b}\overset{(0)}{E_{a}^{~b}}
	-i \omega e_{(2)}^a e_{(3)}^b \bar{\nabla}_{b}\bar{u}^c\overset{(0)}{E_{ac}}
	\nonumber \\
	\!&-&\!
	3i \omega e_{(2)}^a \bar{u}^b\bar{\nabla}_{b}\bar{u}^c\overset{(0)}{E_{ac}}
	+2i \omega e_{(2)}^a e_{(3)}^b \bar{u}^c \bar{\nabla}_c \overset{(0)}{E_{ab}}.
\end{eqnarray}
In the limit $c_1=c_2=c_3=c_4=0$, Einstein–\AE ther theory reduces to general relativity, and the \AE ther field $u^{a}$ becomes an arbitrary unit timelike vector field, unconstrained by the field equations. Therefore, when $c_{13}=0$, the amplitude evolution equation (\ref{tensor part of the next-to-leading-order Tensor modes}) for the tensor modes coincides with that in general relativity. Moreover, for the mixed vector–$x$ and vector–$y$ modes, Eqs. (\ref{P2 for tensor mode}) and (\ref{P3 for tensor mode}) are identical to those in general relativity, since all terms in these equations are independent of the parameters $c_1$, $c_2$, $c_3$ and $c_4$. Consequently, this implies that general relativity and Einstein–\AE ther theory cannot be distinguished through observations of the mixed vector-$x$ and vector-$y$ modes.

For the breathing mode, we find that
\begin{eqnarray}
	\label{P6 for tensor mode}
	2\, P_6\, e^{-i\frac{\theta}{\epsilon}}
	\!&=&\!
	-i \omega e_{(1)}^a e_{(1)}^b \bar{u}^c  \bar{\nabla}_b\overset{(0)}{E_{ac}}
	-i \omega e_{(2)}^a e_{(2)}^b \bar{u}^c  \bar{\nabla}_b\overset{(0)}{E_{ac}}
	\nonumber \\
	\!&+&\!
	i \omega e_{(1)}^a e_{(1)}^b \bar{u}^c  \bar{\nabla}_c\overset{(0)}{E_{ab}}
	+i \omega e_{(2)}^a e_{(2)}^b \bar{u}^c  \bar{\nabla}_c\overset{(0)}{E_{ab}}
	\nonumber \\
	\!&+&\!
	\frac{1}{2}  \frac{c_2-c_{14}}{c_{14}-c_2+2c_2c_{14}}i\omega\bar{\nabla}^{b}\bar{u}^a\overset{(0)}{E_{ab}}.
\end{eqnarray}
Finally, the longitudinal mode satisfies
\begin{eqnarray}
	\label{P1 for tensor mode}
	2\, P_1\, e^{-i\frac{\theta}{\epsilon}}
	\!&=&\!
	i\omega e_{(3)}^a \bar{\nabla}_b \overset{(0)}{E_{a}^{~b}}
    +
   \frac{3}{4} \frac{c_2-c_{14}}{c_{14}-c_2+2c_2c_{14}}i\omega\bar{\nabla}^{b}\bar{u}^a\overset{(0)}{E_{ab}}.
\end{eqnarray}
By contrast, in general relativity the corresponding results are
\begin{eqnarray}
	\label{P6 for tensor mode in GR}
	2\, P_6\, e^{-i\frac{\theta}{\epsilon}}
	\!&=&\!
	-i \omega e_{(1)}^a e_{(1)}^b \bar{u}^c  \bar{\nabla}_b\overset{(0)}{E_{ac}}
	-i \omega e_{(2)}^a e_{(2)}^b \bar{u}^c  \bar{\nabla}_b\overset{(0)}{E_{ac}}
	\nonumber \\
	\!&+&\!
	i \omega e_{(1)}^a e_{(1)}^b \bar{u}^c  \bar{\nabla}_c\overset{(0)}{E_{ab}}
	+i \omega e_{(2)}^a e_{(2)}^b \bar{u}^c  \bar{\nabla}_c\overset{(0)}{E_{ab}}
	\nonumber \\
	\!&-&\!
	\frac{1}{2} i\omega\bar{\nabla}^{b}\bar{u}^a\overset{(0)}{E_{ab}}.
	\\
	\label{P1 for tensor mode in GR}
	2\, P_1\, e^{-i\frac{\theta}{\epsilon}}
	\!&=&\!
	i\omega e_{(3)}^a \bar{\nabla}_b \overset{(0)}{E_{a}^{~b}}
	-
	\frac{3}{4} i\omega\bar{\nabla}^{b}\bar{u}^a\overset{(0)}{E_{ab}}.
\end{eqnarray}
It can be seen that the results for the mixed breathing and longitudinal modes differ from those predicted by general relativity. Therefore, in principle, Einstein-\AE ther theory can be probed by detecting the mixed breathing and longitudinal modes.

\section{Conclusion}
\label{sec: 7}

In this work, we have analyzed the propagation and polarization properties of gravitational waves in Einstein–\AE ther theory on curved backgrounds at both leading and next-to-leading orders within the geometric optics approximation. We have assumed that, within the region of interest, the background \AE ther field is everywhere hypersurface orthogonal. This enables us to adopt the background \AE ther-orthogonal frame, in which the sound cone axis of gravitational waves is everywhere orthogonal to the spatial hypersurfaces. After carrying out the SVT decomposition of the perturbations, this setup ensures that the tensor, vector, and scalar modes decouple at leading order in the geometric optics approximation and leads to particularly simple dispersion relations of gravitational waves. As a result, the analysis is significantly simplified. 

We have therefore analyzed the leading-order geometric optics effects of gravitational waves. At this order, we obtained the propagating degrees of freedom and dispersion relations for the tensor, vector, and scalar modes of gravitational waves, respectively. The results show that gravitational waves possess two independent tensor modes, two independent vector modes, and one independent scalar mode. In the background \AE ther-orthogonal frame, their propagation speeds at the leading order are the same as those in the Minkowski background. We have further derived the gradient stability and no-ghost conditions for these modes at leading order in the geometric optics approximation, which require the theory to satisfy Eq. (\ref{parameters condition}). This extends previous results obtained for the Minkowski \cite{Jacob Oost}, cosmological \cite{Noela Farina Sierra}, and black hole \cite{Shinji Mukohyama1, Antonio De Felice3} backgrounds. Notably, the condition in Eq. (\ref{parameters condition}) coincides with that found in the Minkowski background \cite{Jacob Oost}. This can be attributed to the fact that, in Einstein–\AE ther theory, every term in the field equations involves only two derivative operators. As a result, at leading order in the geometric optics approximation, no terms depending on derivatives of the background arise, and the stability analysis therefore provides no additional information beyond that obtained in Minkowski spacetime. However, when the same approach is applied to Horndeski theory and generalized Proca theory, the stability conditions derived on a curved spacetime background are generically expected to differ from those obtained in Minkowski and cosmological backgrounds.

We have then analyzed the next-to-leading-order effects of the linearized perturbation equations within the geometric optics approximation. At this order, we have derived the amplitude evolution equations for each leading-order mode. For the leading-order tensor modes, we have found that their graviton number is conserved. In contrast, the graviton number for the leading-order vector and scalar modes is not conserved. This indicates that the background is no longer transparent to the vector and scalar modes. In addition to the dispersion/focusing effects induced by the background geometry, the amplitudes of these modes may also experience additional dissipation or amplification due to the interaction with the background fields. Our calculations further indicate that the modes which decouple at leading order mix with each other at next-to-leading order. However, it should be emphasized that the mixing does not change the number of propagating degrees of freedom, and the dispersion relations remain unaffected at this order.  

Subsequently, we have investigated the relationship between these mixed modes and the polarization modes of gravitational waves. In this study, we have assumed that the four-velocity $v^{a}$ of the gravitational wave detector differs only slightly from that of the background \AE ther field $\bar{u}^{a}$. We have found that, when next-to-leading-order effects in the geometric optics approximation are taken into account, mixing occurs between the polarization modes of gravitational waves. Here, we take the case of the leading-order tensor mode as an example—i.e., only $\overset{(0)}{E_{ab}}$ is nonzero in the leading-order amplitude—to further illustrate this point. According to Eq. (\ref{next leading order Eij}), four factors contribute to the mixing of polarization modes. The first factor is that $v^{a}$ is not aligned with $\bar{u}^{a}$, causing the background \AE ther field to acquire spatial components in the proper detector frame, thereby inducing polarization mixing. The second factor is that the next-to-leading order terms of the Riemann curvature tensor still receive contributions from the leading-order amplitude $\overset{(0)}{E_{ab}}$. The third factor is the additional mixed modes, such as $\overset{(1)}{B_{a}}$. Finally, at next-to-leading order, the tensor mode amplitude $\overset{(1)}{E_{ab}}$ is no longer purely transverse, as shown in Eq. (\ref{decomposition next-leading order break transverse conditation}). 

So far, only the $+$ and $\times$ polarization modes of gravitational waves have been detected, and observations indicate that they propagate at the speed of light with extremely high precision. Therefore, we specifically analyzed the polarization mixing effects induced by the leading-order tensor modes in the case $c_{13}=0$. Under this parameter choice, the leading-order tensor modes propagate at the speed of light. We found that the mixed vector-$x$ and vector-$y$ modes cannot be used to distinguish between general relativity and Einstein–\AE ther theory. However, the two theories make different predictions for the mixed breathing and longitudinal modes.

Finally, it should be noted that the present analysis is restricted to the leading and next-to-leading orders in the geometric optics approximation, and the investigation of next-to-next-to-leading order effects remains an interesting avenue for future research. To further illustrate this point, we again take the case of the leading-order tensor mode as an example. In this case, we can solve for the leading-order amplitude $\overset{(0)}{E_{ab}}$ and the additional next-to-leading-order amplitudes, such as $\overset{(1)}{B_{a}}$ and $\overset{(1)}{C}$, only within the leading- and next-to-leading-order approximations of geometric optics. However, the next-to-leading order tensor mode amplitude $\overset{(1)}{E_{ab}}$ remains unknown. Solving for it requires going to the next-to-next-to-leading order. At next-to-next-to-leading order in the geometric optics approximation, the equation describing $\overset{(1)}{E_{ab}}$ is expected to take the form (using an analysis similar to that in Appendix \ref{app: Theoretical framework for analyzing gravitational wave propagation on arbitrary backgrounds})
\begin{eqnarray}
	\label{next-to-next-to-leading order E1}
	\mathcal{F}\big[\overset{(1)}{E_{ab}}\big]
	=
	\mathcal{G}\big[\overset{(0)}{E_{ab}}, \overset{(1)}{B_{a}}, \overset{(1)}{N_{a}}, \overset{(1)}{C},...\big].
\end{eqnarray}
Here, the structure on the left-hand side of the equation is expected to be the same as that of the amplitude evolution equation (\ref{tensor part of the next-to-leading-order Tensor modes}), with $\overset{(0)}{E_{ab}}$ replaced by $\overset{(1)}{E_{ab}}$. The right-hand side of the equation, in contrast, ultimately depends on $\overset{(0)}{E_{ab}}$ and the background spacetime. The above equation has three important features. First, in order for each term in the equation to contribute at the same order in the next-to-next-to-leading approximation, the right-hand side contains an extra factor of $i$ compared with the left-hand side. This causes 
$\overset{(1)}{E_{ab}}$ to have a different phase from $\overset{(0)}{E_{ab}}$, indicating that the next-to-next-to-leading order modifies the phase of the $+$ and $\times$ mode gravitational waves. Second, since the equation depends on the wave vector, $\overset{(1)}{E_{ab}}$ is frequency-dependent. Finally, the left-hand side of Eq. (\ref{next-to-next-to-leading order E1}) is no longer purely algebraic, meaning that solving for $\overset{(1)}{E_{ab}}$ requires integration along the propagation history. Consequently, $\overset{(1)}{E_{ab}}$ encodes information accumulated throughout the wave’s propagation. In different modified gravity theories, the form of Eq. (\ref{next-to-next-to-leading order E1}) varies. Therefore, it is expected that the effects induced by $\overset{(1)}{E_{ab}}$ could be used to test gravity theories and probe the inhomogeneities of matter in the universe. For these reasons, it is necessary to carry out studies of the next-to-next-to-leading-order expansion in the geometric optics approximation in future work.

\section*{Acknowledgments}
YD thanks the Yukawa Institute for Theoretical Physics in Kyoto for their warm hospitality, where this work was carried out and completed. The work of YD and YL is supported in part by the National Natural Science Foundation of China (Grants No. 123B2074, No. 12475056, and No. 12247101), Gansu Province's Top Leading Talent Support Plan, the Fundamental Research Funds for the Central Universities (Grant No. lzujbky-2025-jdzx07), the Natural Science Foundation of Gansu Province (No. 22JR5RA389 and No.25JRRA799), and the `111 Center’ under Grant No. B20063. The work of SM was supported in part by Japan Society for the
Promotion of Science Grants-in-Aid for Scientific Research No. 24K07017 and the World
Premier International Research Center Initiative (WPI), MEXT, Japan.

\appendix

\section{Theoretical framework for analyzing gravitational wave propagation on curved backgrounds}
\label{app: Theoretical framework for analyzing gravitational wave propagation on arbitrary backgrounds}

In this section, we introduce  a general scheme for analyzing the polarization and propagation of gravitational waves in modified gravity theories on a curved spacetime background. This scheme was first introduced in Ref. \cite{Kei-ichiro Kubota}. For notational simplicity, we take a vector-tensor theory with one additional vector field $u^{a}$ as an illustrative example, for which the action can always be written as
\begin{eqnarray}
	\label{action of vector-tensor theory}
	S=\int d^{4}x~\mathcal{L}\left[g_{ab},u^{a}\right].
\end{eqnarray}	
We emphasize that choosing a vector-tensor theory for demonstration does not compromise the generality of the framework. It is chosen solely to make the index structure easier to display. One can readily apply the same procedure directly to analyses of other theories, such as scalar-tensor or tensor-tensor theories. By varying the action separately with respect to $g^{ab}$ and $u^{a}$ respectively, one obtains the corresponding field equations 
\begin{eqnarray}
	\label{field equations of vector-tensor theory M}
	\mathcal{M}_{ab}\left[g_{ab}, u^{a}\right]\!&=&\!0,\\
	\label{field equations of vector-tensor theory N}
	\mathcal{N}_{a}\left[g_{ab}, u^{a}\right]\!&=&\!0.
\end{eqnarray}	

The Isaacson picture \cite{Isaacson1,Isaacson2} indicates that gravitational waves can be well defined in spacetimes where the high- and low-frequency components of the gravitational field can be clearly separated. In this case, $g_{ab}$ and $u^{a}$ can be decomposed into low-frequency background components and high-frequency gravitational wave perturbations:
\begin{eqnarray}
	\label{g=g+h,A=A+a of vector-tensor theory}
	g_{ab}=\bar{g}_{ab}+h_{ab},~ u^{a}=\bar{u}^{a}+w^{a}
\end{eqnarray}	
with the requirement that $\left| h_{ab} \right| \sim \left| w^{a} \right| \sim h \ll 1$. Here, we adopt a coordinate system in which $\bar{g}_{ab}\sim\mathcal{O}(1)$ in the spacetime region of interest, and rescale $\bar{u}^{a}\sim\mathcal{O}(1)$ as well. Accordingly, the field equations (\ref{field equations of vector-tensor theory M}) and (\ref{field equations of vector-tensor theory N}) can be expanded in terms of high-frequency perturbations as
\begin{eqnarray}
	\label{field equations of vector-tensor theory M expanded in terms of high-frequency perturbations}
	\mathcal{M}^{(0)}_{ab}\left[\bar{g}_{ab}, \bar{u}^{a}\right]
	+\mathcal{M}^{(1)}_{ab}\left[\bar{g}_{ab}, \bar{u}^{a}; h_{ab}, w^{a}\right]
	+\mathcal{M}^{(2)}_{ab}\left[\bar{g}_{ab}, \bar{u}^{a}; h_{ab}, w^{a}\right]
	+...\!&=&\!0,\\
	\label{field equations of vector-tensor theory N expanded in terms of high-frequency perturbations}
	\mathcal{N}^{(0)}_{a}\left[\bar{g}_{ab}, \bar{u}^{a}\right]
	+\mathcal{N}^{(1)}_{a}\left[\bar{g}_{ab}, \bar{u}^{a}; h_{ab}, w^{a}\right]
	+\mathcal{N}^{(2)}_{a}\left[\bar{g}_{ab}, \bar{u}^{a}; h_{ab}, w^{a}\right]
	+...\!&=&\!0.
\end{eqnarray}	
The Isaacson picture indicates that the high-frequency components of the field equations govern the propagation of gravitational waves on the background. Consequently, at leading order, the propagation of gravitational waves is naturally described by the linearized perturbation equations \cite{Y.Dong3,Jann Zosso}:
\begin{eqnarray}
	\label{linearized perturbation equation M}
	\mathcal{M}^{(1)}_{ab}\left[\bar{g}_{ab}, \bar{u}^{a}; h_{ab}, w^{a}\right]\!&=&\!0,\\
	\label{linearized perturbation equation N}
	\mathcal{N}^{(1)}_{a}\left[\bar{g}_{ab}, \bar{u}^{a}; h_{ab}, w^{a}\right]\!&=&\!0.
\end{eqnarray}	
The above equations constitute the fundamental equations required to study the propagation and polarization of gravitational waves on curved backgrounds.

\subsection{The leading order of the geometric optics approximation}
\label{sec: A1}

We analyze the linear perturbation equations at the leading order of the geometric optics approximation, subject to the gauge condition (\ref{gauge condition}). Before the analysis, it is useful to outline the structure of Eqs. (\ref{linearized perturbation equation M}) and (\ref{linearized perturbation equation N}). We may formally treat $\bar{g}_{ab}, \bar{u}^{a}$, $h_{ab}$ and $w^{a}$ as tensors defined on a Riemannian spacetime with metric $\bar{g}_{ab}$. By general covariance, Eqs. (\ref{field equations of vector-tensor theory M expanded in terms of high-frequency perturbations}) and (\ref{field equations of vector-tensor theory N expanded in terms of high-frequency perturbations}) remain invariant under coordinate transformations of these quantities. Since tensorial transformations are linear, this invariance holds order by order in the perturbative expansion. Consequently, each term in $\mathcal{M}^{(I)}_{ab}$ and 
$\mathcal{N}^{(I)}_{a}$ ($I=0,1,2,...$) can be written as a combination of $\bar{g}_{ab}, \bar{g}^{ab}, \bar{R}_{abcd}, \bar{u}^{a}, h_{ab}, w^{a}$, and the covariant derivative $\bar{\nabla}_{a}$.

Focusing on the leading order in the geometric optics approximation, we denote the equations as
\begin{eqnarray}
	\label{leading-order equations in the geometric optics approximation}
	\mathcal{M}^{(1,1)}_{ab}\Big[\overset{(0)}{E_{ab}}, \overset{(0)}{B_{a}}, \overset{(0)}{N^{a}}, \overset{(0)}{L}, \overset{(0)}{M}, \overset{(0)}{A}, \overset{(0)}{C}\Big]\!&=&\!0\nonumber\\
	\mathcal{N}^{(1,1)}_{a}\Big[\overset{(0)}{E_{ab}}, \overset{(0)}{B_{a}}, \overset{(0)}{N^{a}}, \overset{(0)}{L}, \overset{(0)}{M}, \overset{(0)}{A}, \overset{(0)}{C}\Big]\!&=&\!0.
\end{eqnarray}
Here, all $\bar{\nabla}_{a}$ acting on the perturbations operate only on the phase and can therefore be replaced by $i k_{a}$. Thus, each term in $\mathcal{M}^{(1,1)}_{ab}$ and $\mathcal{N}^{(1,1)}_{a}$ can be written as a combination of background-dependent parameters, $k_{a}$, and the perturbations. In this sense, $k_{a}$ effectively plays the role of a derivative. Furthermore, substituting the geometric optics expansion (\ref{geometric optics approximation for SVT perturbations}) into Eq. (\ref{transverse and traceless conditation}) shows that the leading-order spatial vectors remain transverse, while the spatial tensor remain transverse and traceless:
\begin{eqnarray}
	\label{transverse and traceless conditation of leading order}
	{}^{(3)}k_{a}\overset{(0)}{N^{a}}={}^{(3)}k^{a}\overset{(0)}{B_{a}}={}^{(3)}k^{a}\overset{(0)}{E_{ab}}=\bar{\gamma}^{ab}\overset{(0)}{E_{ab}}=0.
\end{eqnarray}
Together with the commutativity of $k_{a}$ and the background-dependent parameters, this ensures that the leading-order equations (\ref{leading-order equations in the geometric optics approximation}) retain the same algebraic structure as the linearized perturbation equations on the Minkowski background.

Thus, after projecting $\mathcal{M}^{(1,1)}_{ab}$ and $\mathcal{N}^{(1,1)}_{a}$ along the temporal and spatial directions,
\begin{eqnarray}
	\label{projecting leading order M11,N11 along temporal and spatial directions}
	\mathcal{M}^{(1,1)}_{ab}\!&=&\!
	\mathcal{M}^{(1,1)}_{\perp\perp}\bar{n}_{a}\bar{n}_{b}
	+\mathcal{M}^{(1,1)}_{\parallel\perp a}\bar{n}_{b}+\mathcal{M}^{(1,1)}_{(st)b}\bar{n}_{a}
	+\mathcal{M}^{(1,1)}_{\parallel\parallel ab},
	\nonumber \\
	\mathcal{N}^{(1,1)}_{a}\!&=&\!
	\mathcal{N}^{(1,1)}_{\perp}\bar{n}_{a}
	+\mathcal{N}^{(1,1)}_{\parallel a},
\end{eqnarray}
one can see—by directly applying the corresponding results from the Minkowski background case—that the spatial tensor, vectors, and scalars can be decoupled under the decomposition
\begin{eqnarray}
	\label{decouple under the decomposition leading order}
	\mathcal{M}^{(1,1)}_{\parallel\perp a}\!&=&\!{}^{(V)}\mathcal{M}^{(1,1)}_{\parallel\perp a}+{}^{(3)}k_{a}\,{}^{(S)}\mathcal{M}^{(1,1)}_{\parallel\perp},
	\quad
	\mathcal{N}^{(1,1)}_{\parallel a}={}^{(V)}\mathcal{N}^{(1,1)}_{\parallel a}+{}^{(3)}k_{a}\,{}^{(S)}\mathcal{N}^{(1,1)}_{\parallel},
	\nonumber
	\\
	\mathcal{M}^{(1,1)}_{\parallel\parallel ab}\!&=&\!{}^{(T)}\mathcal{M}^{(1,1)}_{\parallel\parallel ab}+{}^{(3)}k_{a}\,{}^{(V)}\mathcal{M}^{(1,1)}_{\parallel\parallel b}+{}^{(3)}k_{b}\,{}^{(V)}\mathcal{M}^{(1,1)}_{\parallel\parallel a}
	\\
	\nonumber
	\!&+&\!{}^{(3)}k_{a}{}^{(3)}k_{b}\,{}^{(S)}\mathcal{M}^{(1,1)}_{\parallel\parallel}
	+\bar{\gamma}_{ab}{}^{(S)}\tilde{\mathcal{M}}^{(1,1)}_{\parallel\parallel},
\end{eqnarray}
with
\begin{eqnarray}
	\label{decouple under the decomposition leading order transverse and traceless conditation}
	&{}^{(3)}k^{a}\,{}^{(V)}\mathcal{M}^{(1,1)}_{\parallel\perp a}
	={}^{(3)}k^{a}\,{}^{(V)}\mathcal{N}^{(1,1)}_{\parallel a}
	={}^{(3)}k^{a}\,{}^{(V)}\mathcal{M}^{(1,1)}_{\parallel\parallel a}
	=0,\nonumber
	\\
	&{}^{(3)}k^{a}\,{}^{(T)}\mathcal{M}^{(1,1)}_{\parallel\parallel ab}
	=\bar{\gamma}^{ab}\,{}^{(T)}\mathcal{M}^{(1,1)}_{\parallel\parallel ab}
	=0,
\end{eqnarray}
provided that $\bar{u}^{a}$ satisfies $\bar{\gamma}_{ab}\bar{u}^{b}=0$. The decomposition (\ref{decouple under the decomposition leading order}) is unique.

Hence, the tensor mode gravitational waves satisfy the algebraic equation 
\begin{eqnarray}
	\label{tensor mode equation leading order}
	{}^{(T)}\mathcal{M}^{(1,1)}_{\parallel\parallel ab}
	\Big[\overset{(0)}{E_{ab}}\Big]=0.
\end{eqnarray}
Using standard methods for solving linear systems, we can derive the condition that $k_{a}$ must satisfy for the equation to admit nontrivial solutions. This allows us to determine whether the theory supports tensor modes and their corresponding propagation speeds. Similarly, the vector mode gravitational waves satisfy the following equations:
\begin{eqnarray}
	\label{vector mode equation leading order}
	{}^{(V)}\mathcal{N}^{(1,1)}_{\parallel a}
	\Big[\overset{(0)}{B_{a}}, \overset{(0)}{N^{a}}\Big]=0,\quad
	{}^{(V)}\mathcal{M}^{(1,1)}_{\parallel\perp a}
	\Big[\overset{(0)}{B_{a}}, \overset{(0)}{N^{a}}\Big]=0,\quad
	{}^{(V)}\mathcal{M}^{(1,1)}_{\parallel\parallel a}	
	\Big[\overset{(0)}{B_{a}}, \overset{(0)}{N^{a}}\Big]=0.
\end{eqnarray}
Due to general covariance, when the background values $\bar{g}_{ab}$ and $\bar{u}^{a}$ satisfy the field equations, only two of these equations are independent, matching the number of variables. This allows one to determine the existence of vector modes and their wave speeds. There are six algebraic equations describing the scalar modes:
\begin{eqnarray}
	\label{scalar mode equation leading order}
	\mathcal{N}^{(1,1)}_{\perp}
	\Big[\overset{(0)}{L}, \overset{(0)}{M}, \overset{(0)}{A}, \overset{(0)}{C}\Big]\!&=&\!0,\quad
	{}^{(S)}\mathcal{N}^{(1,1)}_{\parallel}
	\Big[\overset{(0)}{L}, \overset{(0)}{M}, \overset{(0)}{A}, \overset{(0)}{C}\Big]=0,
	\nonumber \\
	\mathcal{M}^{(1,1)}_{\perp\perp}
	\Big[\overset{(0)}{L}, \overset{(0)}{M}, \overset{(0)}{A}, \overset{(0)}{C}\Big]\!&=&\!0,\quad
	{}^{(S)}\mathcal{M}^{(1,1)}_{\parallel\perp}
	\Big[\overset{(0)}{L}, \overset{(0)}{M}, \overset{(0)}{A}, \overset{(0)}{C}\Big]=0,
	\nonumber \\
	{}^{(S)}\mathcal{M}^{(1,1)}_{\parallel\parallel}
	\Big[\overset{(0)}{L}, \overset{(0)}{M}, \overset{(0)}{A}, \overset{(0)}{C}\Big]\!&=&\!0,\quad
	{}^{(S)}\tilde{\mathcal{M}}^{(1,1)}_{\parallel\parallel}
	\Big[\overset{(0)}{L}, \overset{(0)}{M}, \overset{(0)}{A}, \overset{(0)}{C}\Big]=0.
\end{eqnarray}
Similarly, only four of these equations are independent, equal to the number of variables. This enables the determination of the existence of scalar modes and their wave speeds. Once the tensor, vector, and scalar modes of the theory are identified, any gravitational wave solution can be represented as a linear combination of these modes, owing to the linearity of the equations.

It should be noted that the leading-order equations (\ref{tensor mode equation leading order})-(\ref{scalar mode equation leading order}) are homogeneous in $k_{a}$. As a result, no graviton mass term appears in the expressions for the wave speeds, i.e., in the gravitational wave dispersion relations. To accommodate cases with massive gravitons (with sufficiently large mass), one would need to artificially adjust the relative scaling between the background-dependent parameters and $\epsilon$ in the equations, allowing zeroth-order terms in $k_{a}$ to appear in the leading-order equations. 

Finally, we note that if the background-dependent parameters in Eqs. (\ref{tensor mode equation leading order})-(\ref{scalar mode equation leading order}) are independent of $\bar{\nabla}_{a}$ and $\bar{R}_{abcd}$, the equations, in essence, provide no information beyond that of the Minkowski case. Hence, the propagation speed of gravitational waves in any spacetime equals that in Minkowski spacetime. General relativity and Einstein–\AE ther theory, which we shall discuss, provide two examples. Their field equations are second-order, with each term containing exactly two derivatives. As a result, at the leading order of the geometric optics approximation, all derivatives act solely on the phase, and the equation parameters are independent of $\bar{\nabla}_{a}$ and $\bar{R}_{abcd}$. In contrast, for Horndeski and generalized Proca theories, although the field equations remain second-order, some terms contain more than two derivatives. At the leading order of the geometric optics approximation, this allows background-dependent parameters to involve derivatives of the background. Consequently, the propagation speed of gravitational waves differs from that in the Minkowski background.

\subsection{The next-to-leading order of the geometric optics approximation}
\label{sec: A2}
We now consider Eqs. (\ref{linearized perturbation equation M}) and (\ref{linearized perturbation equation N}) at the next-to-leading order in the geometric optics approximation: $\mathcal{M}^{(1,2)}_{ab}=0$, $\mathcal{N}^{(1,2)}_{a}=0$. In these equations, the next-to-leading-order amplitude terms in the geometric optics expansion (\ref{geometric optics approximation for SVT perturbations}), such as $\overset{(1)}{E_{ab}}$, appear only in terms where all derivatives acting on the perturbations operate on the phase. This implies that $\mathcal{M}^{(1,2)}_{ab}=0$ and $\mathcal{N}^{(1,2)}_{a}=0$ hcan be equivalently written as:
\begin{eqnarray}
	\label{relation between leading and next leading order of M2}
	\mathcal{M}^{(1,1)}_{ab}\Big[\overset{(1)}{E_{ab}}, \overset{(1)}{B_{a}}, \overset{(1)}{N^{a}}, \overset{(1)}{L}, \overset{(1)}{M}, \overset{(1)}{A}, \overset{(1)}{C}\Big]
	\!&=&\!
	\mathcal{F}_{ab}\Big[\overset{(0)}{E_{ab}}, \overset{(0)}{B_{a}}, \overset{(0)}{N^{a}}, \overset{(0)}{L}, \overset{(0)}{M}, \overset{(0)}{A}, \overset{(0)}{C}\Big],
	\\
	\label{relation between leading and next leading order of N2}
	\mathcal{N}^{(1,1)}_{a}\Big[\overset{(1)}{E_{ab}}, \overset{(1)}{B_{a}}, \overset{(1)}{N^{a}}, \overset{(1)}{L}, \overset{(1)}{M}, \overset{(1)}{A}, \overset{(1)}{C}\Big]
	\!&=&\!
	\mathcal{G}_{a}\Big[\overset{(0)}{E_{ab}}, \overset{(0)}{B_{a}}, \overset{(0)}{N^{a}}, \overset{(0)}{L}, \overset{(0)}{M}, \overset{(0)}{A}, \overset{(0)}{C}\Big].
\end{eqnarray}

Similarly, Eqs. (\ref{relation between leading and next leading order of M2}) and (\ref{relation between leading and next leading order of N2}) can be decomposed by following the same procedure as used for Eqs. (\ref{projecting leading order M11,N11 along temporal and spatial directions}) and (\ref{decouple under the decomposition leading order}). At this stage, the spatial tensor, vectors, and scalars are generally no longer decoupled in the decomposition of the right-hand side of the equations, since the derivatives acting on the perturbations in $\mathcal{F}_{ab}$ and $\mathcal{G}_{a}$ no longer act exclusively on the phase. For the terms on the left-hand side, structurally, they merely replace all zeroth-order amplitude variables in Eq. (\ref{leading-order equations in the geometric optics approximation}) with first-order ones. However, these first-order amplitudes generally do not satisfy the transverse conditions. Consequently, in the decomposition of the left-hand side, the spatial tensor, vectors, and scalars are generally not decoupled. Nevertheless, the terms that cause this coupling always involve contractions of the spatial tensor and vectors with ${}^{(3)}k_{a}$. By substituting Eq. (\ref{geometric optics approximation for SVT perturbations}) into the transverse condition (\ref{transverse and traceless conditation}) and considering the next-to-leading order, one finds that
\begin{eqnarray}
	\label{decomposition next-leading order break transverse conditation}
	i\epsilon\,{}^{(3)}k_{a}\,\overset{(1)}{B^{a}}
	\!&=&\!-
	\bar{\gamma}^{ab}\bar{\nabla}_{a}\overset{(0)}{B_{b}},\nonumber
	\\
	i\epsilon\,{}^{(3)}k_{a}\,\overset{(1)}{N^{a}}
	\!&=&\!-
	\bar{\gamma}^{ab}\bar{\nabla}_{a}\overset{(0)}{N_{b}},
	\\
	i\epsilon\,{}^{(3)}k_{a}\overset{(1)}{E^{ab}}
	\!&=&\!-
	\bar{\gamma}^{cd}\bar{\gamma}^{be}\bar{\nabla}_{d}\overset{(0)}{E_{ce}}.\nonumber
\end{eqnarray}
This shows that these terms can be expressed in terms of zeroth-order amplitude variables, allowing them to be moved to the right-hand side and absorbed into $\mathcal{F}_{ab}$ and $\mathcal{G}_{a}$. As a result, the left-hand side can then be fully decoupled.

Thus, at the next-to-leading order, the equations describing the tensor modes necessarily take the following form:
\begin{eqnarray}
	\label{tensor mode equation next leading order}
	{}^{(T)}{\mathcal{M}}^{(1,1)}_{\parallel\parallel ab}
	\Big[\overset{(1)}{E_{ab}}\Big]
	=
	{}^{(T)}\mathcal{F}_{ab}\Big[\overset{(0)}{E_{ab}}, \overset{(0)}{B_{a}}, \overset{(0)}{N^{a}}, \overset{(0)}{L}, \overset{(0)}{M}, \overset{(0)}{A}, \overset{(0)}{C}\Big].
\end{eqnarray}
Here, ${}^{(T)}\mathcal{F}_{ab}$ depends solely on the zeroth-order quantities of the perturbations in the geometric optics approximation. In addition, we need to clarify the potential ambiguity associated with ${}^{(T)}{\mathcal{M}}^{(1,1)}_{\parallel\parallel ab}\Big[\overset{(1)}{E_{ab}}\Big]$. In Sec. \ref{sec: A1}, two definitions of the mapping ${}^{(T)}{\mathcal{M}}^{(1,1)}_{\parallel\parallel ab}$ are actually given. The first, in Eq. (\ref{decouple under the decomposition leading order}), refers to the quantity being decomposed, while the second, in Eq. (\ref{tensor mode equation leading order}), refers to the left-hand side of the decoupled equations. For zeroth-order amplitudes, the two definitions are equivalent, since they satisfy the transverse condition in Eq. (\ref{decouple under the decomposition leading order transverse and traceless conditation}). However, first-order amplitudes are generally not transverse, so the definitions are no longer equivalent. As discussed above, the non-decoupled terms can be moved to the right-hand side using Eq. (\ref{tensor mode equation next leading order}). Therefore, we adopt the latter definition of 
${}^{(T)}{\mathcal{M}}^{(1,1)}_{(ss)ab}$ here. For the subsequent discussion of the vector and scalar modes, the quantities are defined in the same manner.

For vector modes, the corresponding equations are
\begin{eqnarray}
	\label{vector mode equation next leading order}
	{}^{(V)}\mathcal{N}^{(1,1)}_{\parallel a}
	\Big[\overset{(1)}{B_{a}}, \overset{(1)}{N^{a}}\Big]
	\!&=&\!
	{}^{(V)}\mathcal{F}^{1}_{a}\Big[\overset{(0)}{E_{ab}}, \overset{(0)}{B_{a}}, \overset{(0)}{N^{a}}, \overset{(0)}{L}, \overset{(0)}{M}, \overset{(0)}{A}, \overset{(0)}{C}\Big],
	\nonumber \\
	{}^{(V)}\mathcal{M}^{(1,1)}_{\parallel\perp a}
	\Big[\overset{(1)}{B_{a}}, \overset{(1)}{N^{a}}\Big]
	\!&=&\!
	{}^{(V)}\mathcal{F}^{2}_{a}\Big[\overset{(0)}{E_{ab}}, \overset{(0)}{B_{a}}, \overset{(0)}{N^{a}}, \overset{(0)}{L}, \overset{(0)}{M}, \overset{(0)}{A}, \overset{(0)}{C}\Big],
	\nonumber \\
	{}^{(V)}\mathcal{M}^{(1,1)}_{\parallel\parallel a}	
	\Big[\overset{(1)}{B_{a}}, \overset{(1)}{N^{a}}\Big]
	\!&=&\!
	{}^{(V)}\mathcal{F}^{3}_{a}\Big[\overset{(0)}{E_{ab}}, \overset{(0)}{B_{a}}, \overset{(0)}{N^{a}}, \overset{(0)}{L}, \overset{(0)}{M}, \overset{(0)}{A}, \overset{(0)}{C}\Big].
\end{eqnarray}
Owing to general covariance, when the background satisfies the field equations and the leading-order equations (\ref{tensor mode equation leading order})-(\ref{scalar mode equation leading order}) hold, only two of the three equations above are independent. For scalar modes, the equations are
\begin{eqnarray}
	\label{scalar mode equation next leading order}
	\mathcal{N}^{(1,1)}_{\perp}
	\Big[\overset{(1)}{L}, \overset{(1)}{M}, \overset{(1)}{A}, \overset{(1)}{C}\Big]
	\!&=&\!
	{}^{(S)}\mathcal{F}^{1}\Big[\overset{(0)}{E_{ab}}, \overset{(0)}{B_{a}}, \overset{(0)}{N^{a}}, \overset{(0)}{L}, \overset{(0)}{M}, \overset{(0)}{A}, \overset{(0)}{C}\Big],
	\nonumber \\
	{}^{(S)}\mathcal{N}^{(1,1)}_{\parallel}
	\Big[\overset{(1)}{L}, \overset{(1)}{M}, \overset{(1)}{A}, \overset{(1)}{C}\Big]	
	\!&=&\!
	{}^{(S)}\mathcal{F}^{2}\Big[\overset{(0)}{E_{ab}}, \overset{(0)}{B_{a}}, \overset{(0)}{N^{a}}, \overset{(0)}{L}, \overset{(0)}{M}, \overset{(0)}{A}, \overset{(0)}{C}\Big],
	\nonumber \\
	\mathcal{M}^{(1,1)}_{\perp\perp}
	\Big[\overset{(1)}{L}, \overset{(1)}{M}, \overset{(1)}{A}, \overset{(1)}{C}\Big]
	\!&=&\!
	{}^{(S)}\mathcal{F}^{3}\Big[\overset{(0)}{E_{ab}}, \overset{(0)}{B_{a}}, \overset{(0)}{N^{a}}, \overset{(0)}{L}, \overset{(0)}{M}, \overset{(0)}{A}, \overset{(0)}{C}\Big],
	\nonumber \\
	{}^{(S)}\mathcal{M}^{(1,1)}_{\parallel\perp}
	\Big[\overset{(1)}{L}, \overset{(1)}{M}, \overset{(1)}{A}, \overset{(1)}{C}\Big]
	\!&=&\!
	{}^{(S)}\mathcal{F}^{4}\Big[\overset{(0)}{E_{ab}}, \overset{(0)}{B_{a}}, \overset{(0)}{N^{a}}, \overset{(0)}{L}, \overset{(0)}{M}, \overset{(0)}{A}, \overset{(0)}{C}\Big],
	\nonumber \\
	{}^{(S)}\mathcal{M}^{(1,1)}_{\parallel\parallel}
	\Big[\overset{(1)}{L}, \overset{(1)}{M}, \overset{(1)}{A}, \overset{(1)}{C}\Big]
	\!&=&\!
	{}^{(S)}\mathcal{F}^{5}\Big[\overset{(0)}{E_{ab}}, \overset{(0)}{B_{a}}, \overset{(0)}{N^{a}}, \overset{(0)}{L}, \overset{(0)}{M}, \overset{(0)}{A}, \overset{(0)}{C}\Big],
	\nonumber \\
	{}^{(S)}\tilde{\mathcal{M}}^{(1,1)}_{\parallel\parallel}
	\Big[\overset{(1)}{L}, \overset{(1)}{M}, \overset{(1)}{A}, \overset{(1)}{C}\Big]
	\!&=&\!
	{}^{(S)}\mathcal{F}^{6}\Big[\overset{(0)}{E_{ab}}, \overset{(0)}{B_{a}}, \overset{(0)}{N^{a}}, \overset{(0)}{L}, \overset{(0)}{M}, \overset{(0)}{A}, \overset{(0)}{C}\Big].
\end{eqnarray}
Similarly, when the background satisfies the field equations and Eqs. (\ref{tensor mode equation leading order})-(\ref{scalar mode equation leading order}) hold, only four of the above equations are independent.

We now examine the next-to-leading-order effects of the leading-order tensor modes, considering the case where only $\overset{(0)}{E_{ab}}$ is nonzero among the zeroth-order amplitudes. Since 
$\overset{(0)}{E_{ab}}$ satisfies Eq. (\ref{tensor mode equation leading order}), the left-hand side of Eq. (\ref{tensor mode equation next leading order}) vanishes. This results in an equation involving only $\overset{(0)}{E_{ab}}$, ${}^{(T)}\mathcal{F}_{ab}\Big[\overset{(0)}{E_{ab}}\Big]=0$, which governs the evolution of the amplitude of tensor modes on any background. For Eqs. (\ref{vector mode equation next leading order}) and (\ref{scalar mode equation next leading order}), the left-hand sides are algebraic equations for the spatial vectors and scalars, respectively, while the right-hand sides are source terms depending solely on $\overset{(0)}{E_{ab}}$. This directly determines the amplitudes of the mixed vector and scalar modes associated with the tensor mode as it propagates on a curved spacetime background. At the next-to-leading-order level, the dispersion relations of these mixed modes are the same as those of the leading-order tensor mode. By a similar analysis, one can derive the evolution equations for the amplitudes of the leading-order vector and scalar modes from Eqs. (\ref{vector mode equation next leading order})-(\ref{scalar mode equation next leading order}), and determine the corresponding mixed tensor, vector, and scalar mode contributions. After analyzing the cases where the leading-order modes are purely tensor, vector, or scalar, the next-to-leading-order effects of an arbitrary gravitational wave can be expressed as a linear combination of these results.

\section{Next-to-leading-order equations of vector and scalar modes}
\label{app: Next-to-leading-order equations of vector and scalar modes}
\subsection{Vector modes}
\label{Vector modes}

The amplitude evolution equation for the vector modes is obtained as
\begin{eqnarray}
	\label{vector part of the next-to-leading-order Vector modes}
	&-8v_{T}^{-6}v_{V}^4c_{14}^2\bar{\nabla}_{a}{}^{(3)}k^{a}\overset{(0)}{N_{g}}
	-16v_{T}^{-6}v_{V}^4c_{14}^2{}^{(3)}k^{a}\bar{\nabla}_{a}\overset{(0)}{N_{g}}
	\nonumber \\
	&-8v_{T}^{-6}v_{V}^2
	c_{14}^2
	\bar{\nabla}_{a}\left(\omega\bar{u}^{a}\right)\overset{(0)}{N_{g}}
	-16v_{T}^{-6}v_{V}^2
	c_{14}^2
	\omega\bar{u}^{a}\bar{\nabla}_{a}\overset{(0)}{N_{g}}
	\nonumber\\
	&-\frac{8c_{14}^2v_{T}^{-6}v_{V}^6}{\omega}
	{}^{(3)}k^{a}k_{g}\bar{u}^{b}\bar{\nabla}_{a}\overset{(0)}{N_{b}}
	+\frac{8c_{14}^2v_{T}^{-6}v_{V}^6}{\omega}
	{}^{(3)}k^{a}\bar{u}_{g}\bar{u}^{b}\bar{\nabla}_{a}\overset{(0)}{N_{b}}
	\nonumber \\
	&+\frac{4c_{14}v_{T}^{-4}v_{V}^4\left[2c_4+c_{13}\left(3c_1-c_3+2c_4\right)\right]}{\omega}{}^{(3)}k^{a}{}^{(3)}k_{g}\bar{\nabla}_{a}\bar{u}_{b}\overset{(0)}{N^{b}}
	\nonumber \\
	&-{4c_{14}v_{T}^{-4}v_{V}^4\left(-4c_1+7c_1^2+6c_1c_3-c_3^2-2c_4+6c_{13}c_4\right)}{}^{(3)}k^{a}\bar{u}_{g}\bar{\nabla}_{a}\bar{u}_{b}\overset{(0)}{N^{b}}
	\nonumber \\
	&+8\left(c_1-c_3\right)v_{T}^{-6}v_{V}^2c_{14}^2\omega\bar{\nabla}_{a}\bar{u}_{g}\overset{(0)}{N^{a}}
	\nonumber \\
	&-8\left(c_1-c_3\right)v_{T}^{-6}v_{V}^2c_{14}^2\omega\bar{u}^{b}\bar{u}_{g}\bar{\nabla}_{b}\bar{u}_{a}\overset{(0)}{N^{a}}
	-8\left(c_1-c_3\right)v_{T}^{-6}v_{V}^2c_{14}^2\omega\bar{\nabla}_{g}\bar{u}_{a}\overset{(0)}{N^{a}}
	\nonumber \\
	&+4\left[2c_1^3+c_3^2\left(1-2c_4\right)-2c_4+c_1^2\left(-5+2c_4\right)-2c_1\left(c_3^2+2c_4\right)\right]
	v_{T}^{-4}v_{V}^2c_{14}\omega\bar{\nabla}_{a}\bar{u}_{g}\overset{(0)}{N^{a}}
	\nonumber \\
	&-4\left[2c_1^3+c_3^2\left(1-2c_4\right)-2c_4+c_1^2\left(-5+2c_4\right)-2c_1\left(c_3^2+2c_4\right)\right]v_{T}^{-4}v_{V}^2c_{14}\omega\bar{\nabla}_{g}\bar{u}_{a}\overset{(0)}{N^{a}}
	\nonumber \\
	&-4v_{T}^{-4}v_{V}^2c_{14}\left[c_1^2+c_3^2-2c_4+2c_3c_4+2c_1c_{34}\right]\bar{u}^{a}\bar{u}^{b}\bar{\nabla}_{b}{}^{(3)}k_{a}\overset{(0)}{N^{g}}
	\nonumber \\
	&+16\frac{v_{T}^{-6}v_{V}^6c_{14}^{2}}{\omega^2}{}^{(3)}k^{a}{}^{(3)}k^{b}{}^{(3)}k_{g}\bar{\nabla}_{b}\overset{(0)}{N_{a}}
	-16\frac{v_{T}^{-6}v_{V}^4c_{14}^2}{\omega}{}^{(3)}k^{a}{}^{(3)}k_{g}\bar{u}^{b}\bar{\nabla}_{b}\overset{(0)}{N_{a}}
	\nonumber \\
	&+16v_{T}^{-6}v_{V}^{4}c_{14}^2{}^{(3)}k_{g}\bar{u}^{a}\bar{u}^{b}\bar{\nabla}_{b}\overset{(0)}{N_{a}}
    -16\frac{v_{T}^{-6}v_{V}^6c_{14}^2}{\omega}{}^{(3)}k^{a}{}^{(3)}k^{b}\bar{u}_{g}\bar{\nabla}_{b}\overset{(0)}{N_{a}}
	\nonumber \\
	&+16v_{T}^{-6}v_{V}^4c_{14}^2{}^{(3)}k^{a}\bar{u}^{b}\bar{u}_{g}\bar{\nabla}_{b}\overset{(0)}{N_{a}}
    +16v_{T}^{-6}v_{V}^4c_{14}^2\omega\bar{u}^{a}\bar{u}^{b}\bar{u}_{g}\bar{\nabla}_{b}\overset{(0)}{N_{a}}
	\nonumber \\
	&-\frac{4v_{T}^{-4}v_{V}^4c_{14}^2\left(2c_4+c_{13}\left(3c_1-c_3+2c_4\right)\right)}{c_{14}\omega}{}^{(3)}k^{a}{}^{(3)}k_{g}\bar{\nabla}_{b}\bar{u}_{a}\overset{(0)}{N^{b}}
	\nonumber \\
	&-\frac{4v_{T}^{-4}v_{V}^4c_{14}^2\left(2c_4+c_{13}\left(3c_1-c_3+2c_4\right)\right)}{c_{14}}{}^{(3)}k_{g}\bar{u}^{b}\bar{\nabla}_{b}\bar{u}_{a}\overset{(0)}{N^{a}}
	\nonumber \\
	&-\frac{8v_{T}^{-4}v_{V}^4c_{14}^2\left[2c_4+c_{13}\left(-3c_1+2c_1^2-c_3+2c_1c_3+2c_4\left(-2+c_{13}\right)\right)\right]}{\left(-2+c_1\right)c_1-c_3^2}{}^{(3)}k^{a}\bar{u}^{b}\bar{\nabla}_{b}\bar{u}_{a}\overset{(0)}{N_{g}}
	\nonumber \\
	&+\frac{4v_{T}^{-4}v_{V}^4c_{14}^2\left[2c_4+c_{13}\left(3c_1-c_3+2c_4\right)\right]}{c_{14}}{}^{(3)}k^{a}\bar{u}_{g}\bar{\nabla}_{b}\bar{u}_{a}\overset{(0)}{N^{b}}
	\nonumber \\
	&-\Big[
	4c_1^5+c_3^4-4c_3^2c_4-4\left(-1+c_3\right)\left(3-2c_3+c_3^2\right)c_4^2+c_1^4\left(-3+4c_3+8c_4\right)
	\nonumber \\
	&-4c_1c_4\left(-4+2c_3\left(5-3c_3+c_3^2\right)+\left(3-2c_3+c_3^2\right)c_4\right)
	\nonumber \\
	&+4c_1^3\left(-c_3^2+2c_3\left(1+c_4\right)+\left(-3+c_4\right)\left(1+c_4\right)\right)
	\nonumber \\
	&-2c_1^2\left(
	-4+2c_3^3-2c_3\left(-1+c_4\right)\left(5+c_4\right)+2c_4\left(5+c_4\right)
	+c_3^2\left(-5+4c_4\right)
	\right)
	\Big]
	\nonumber \\
	&\times \frac{4v_{T}^{-4}v_{V}^4c_{14}^2}{c_{14}\left(\left(-2+c_1\right)c_1-c_3^2\right)}\omega\bar{u}^{b}\bar{u}_{g}\bar{\nabla}_{b}\bar{u}_{a}\overset{(0)}{N^{a}}=0.
\end{eqnarray}

The mixed tensor mode amplitude associated with the leading-order vector mode satisfies
\begin{eqnarray}
	\label{tensor part of the next-to-leading-order Vector modes}
	&\epsilon\overset{(1)}{E_{fg}}=
	-\frac{i}{{}^{(3)}k^{4}+\left(-1+c_{13}\right){}^{(3)}k^{2}\omega^2}
	\Big[
	-2v_{T}^{-2}c_{14}\omega
	\nonumber \\
	&\times \Big(
	{}^{(3)}k^{2}\bar{u}^{a}\bar{\nabla}_{a}\bar{u}_{f}\overset{(0)}{N_{g}}
	+{}^{(3)}k^{2}\bar{u}^{a}\bar{\nabla}_{a}\bar{u}_{g}\overset{(0)}{N_{f}}
	+\left({}^{(3)}k_{f}{}^{(3)}k_{g}-{}^{(3)}k^2\bar{\gamma}_{fg}\right)\bar{u}^{b}\bar{\nabla}_{b}\bar{u}_{a}\overset{(0)}{N^{a}}
	\Big)
	\nonumber \\
	&+{}^{(3)}k^{a}
	\big[
	-2c_{T}^{-2}v_{V}^2c_{14}
    \left({}^{(3)}k_{f}{}^{(3)}k_{g}-{}^{(3)}k^2\bar{\gamma}_{fg}\right)\left(
	\bar{\nabla}_{a}\bar{u}_{b}-
	\bar{\nabla}_{b}\bar{u}_{a}\right)\overset{(0)}{N^{b}}
	\nonumber \\
	&-2c_{T}^{-2}v_{V}^2c_{14}{}^{(3)}k^2\bar{\nabla}_{a}\bar{u}_{f}\overset{(0)}{N_{g}}
	-2c_{T}^{-2}v_{V}^2c_{14}{}^{(3)}k^2\bar{\nabla}_{f}\bar{u}_{a}
	\overset{(0)}{N_{g}}
	\nonumber \\
	&
	+\Big(2v_{T}^{-2}c_{14}\omega{}^{(3)}k_{f}
	+2v_{T}^{-2}v_{V}^2c_{14}{}^{(3)}k^2\bar{u}_{f}\Big)\bar{u}^{b}\bar{\nabla}_{b}\bar{u}_{a}\overset{(0)}{N_{g}}
	\nonumber \\
	&
	-2v_{T}^{-2}v_{V}^2c_{14}{}^{(3)}k^2\bar{\nabla}_{a}\bar{u}_{g}\overset{(0)}{N_{f}}
	+2v_{T}^{-2}v_{V}^2c_{14}{}^{(3)}k^2\bar{\nabla}_{g}\bar{u}_{a}
	\overset{(0)}{N_{f}}
	\nonumber \\
	&
	+\Big(2v_{T}^{-2}c_{14}\omega{}^{(3)}k_{g}
	+2v_{T}^{-2}v_{V}^2c_{14}{}^{(3)}k^2\bar{u}_{g}\Big)\bar{u}^{b}\bar{\nabla}_{b}\bar{u}_{a}\overset{(0)}{N_{f}}
	\big]\Big].
\end{eqnarray}

The mixed scalar mode amplitudes associated with the leading-order vector mode satisfy
\begin{eqnarray}
	\label{scalar part of the next-to-leading-order Vector modes C}
    &\epsilon\overset{(1)}{C}
    =
    \frac{i}{\left(c_1+2c_3-c_4\right){}^{(3)}k^2\omega}
    \nonumber \\
    &\times\Big[
    -\Big(
    c_1^3+c_1^2\left(2+5c_3\right)
    -c_3^2\left(c_3-4c_4\right)
    +c_1\left(3c_3^2+4c_4+c_3\left(-2+4c_4\right)\right)
    \Big)
    \nonumber \\
    &\times\left(\bar{\nabla}_{a}\bar{u}_{b}-\bar{\nabla}_{b}\bar{u}_{a}\right)\omega{}^{(3)}k^{a}\overset{(0)}{N^{b}}
    \nonumber \\
    &+2c_{13}{}^{(3)}k^{a}
    \Big(
    -\left(c_1^2+c_3^2-2c_4+2c_3c_4+2c_1c_{34}\right)\omega\bar{u}^{b}\bar{\nabla}_{b}\overset{(0)}{N_{a}}
    +2v_{T}^{-2}v_{V}^2c_{14}k_{b}\bar{\nabla}^{b}\overset{(0)}{N_{a}}
    \Big)
    \nonumber \\
    &+{}^{(3)}k^2\Big(
    2iv_{T}^{-2}c_{14}\omega
    \epsilon\overset{(1)}{L}
    +\left(c_1^3+c_1^2\left(-2+c_3\right)
    -c_3^3-c_1c_3\left(2+c_3\right)
    \right)
    \nonumber \\
    &
    \times\big(2\bar{u}^{b}\bar{\nabla}_{b}\bar{u}_{a}\overset{(0)}{N^{a}}
    +\bar{\gamma}_{ab}\bar{\nabla}^{b}\overset{(0)}{N^{a}}
    \big)
    \Big)
    \nonumber \\
    &+c_{14}\omega
    \Big(
    -2\left(
    c_1^2
    +(-1+c_3)c_3+c_1(-1+2c_3)
    \right)\omega\bar{\gamma}_{ab}\bar{\nabla}^{b}\overset{(0)}{N^{a}}
    \nonumber \\
    &-
    \left(
    c_1^2+c_3^2-2c_4+2c_3c_4+2c_1c_{34}
    \right)\omega\bar{u}^{b}\bar{\nabla}_{b}\bar{u}_{a}\overset{(0)}{N^{a}}
    \nonumber \\
    &
    -2v_{T}^{-2}v_{V}^2c_{14}k_{b}
    \left(\bar{\nabla}_{a}\bar{u}^{b}-\bar{\nabla}^{b}\bar{u}_{a}\right)\overset{(0)}{N^{a}}
    \Big)
    \Big],
    \\
    \label{scalar part of the next-to-leading-order Vector modes M}
    &\epsilon\overset{(1)}{M}
    =
    \frac{1}{2\left(c_1 + 2c_3 - c_4\right) {}^{(3)}k^2 \omega^2}
    \Big[
    k_{a}\Big(
    -2\big(c_{13}^2 -2v_{T}^{-2}c_4\big)\omega\bar{u}_{b}
    \bar{\nabla}^{b}\overset{(0)}{N^{a}}
    \nonumber \\
    &
    +\big(c_1^2 - 2c_1(3+c_{34}) - c_3(3c_3 + 2c_4)\big) 
    \big(\bar{\nabla}^{a}\bar{u}^{b}
    - \bar{\nabla}^{b}\bar{u}^{a}\big)\omega\overset{(0)}{N_{b}}
    \nonumber \\
    &
    +4v_{T}^{-2}v_{V}^2c_{14}
   k_{b}\bar{\nabla}^{b}\overset{(0)}{N^{a}}
    \Big)
    +{}^{(3)}k^2\Big(
    -2i(-2+c_{14})\omega \epsilon\overset{(1)}{L}
    \nonumber \\
    &
    +-2v_{T}^{-2}v_{V}^2c_{14}
    \big(
    2\bar{u}_{b}\bar{\nabla}^{b}\bar{u}^{a}\overset{(0)}{N_{a}}
    +\bar{\gamma}_{ab}\bar{\nabla}^{b}\overset{(0)}{N^{a}}
    \big)
    \Big)
    \nonumber \\
    &
    +2\omega\Big(
    v_{T}^{-2}c_{14}\omega
    \bar{\gamma}_{ab}\bar{\nabla}^{b}\overset{(0)}{N^{a}}
    +\overset{(0)}{N_{a}}\Big(
    -(c_{13}^2 - 2v_{T}^{-2}c_4)\omega\bar{u}_{b}\bar{\nabla}^{b}\bar{u}^{a}
    \nonumber \\
    &
    -2v_{T}^{-2}v_{V}^2 c_{14}
    k_{b}
    (\bar{\nabla}^{a}\bar{u}^{b}
    - \bar{\nabla}^{b}\bar{u}^{a})
    \Big)
    \Big)
    \Big], 
    \\ 
    \label{scalar part of the next-to-leading-order Vector modes L}
    &\epsilon\overset{(1)}{L}
    = 
    \frac{-i}{2 c_{123}\left(-2+c_{14}\right)
    	{}^{(3)}k^4\omega
    	-2\left(-1+c_{13}\right)\left(2+c_{13}+3c_2\right)
    	c_{14}{}^{(3)}k^2\omega^3}
    \nonumber \\
    &\times\Big[
    \Big(
    (c_1^3 + c_1^2(2- c_2 + 5c_3)
    + c_3(3c_2c_3 - c_3^2 + 2c_2c_4 + 4 c_3 c_4)
    \nonumber \\
    &
    + c_1(c_3(-2+3c_3)+4(1+c_3)c_4+2c_2(3+c_{34})))
    {}^{(3)}k^2\omega
    \nonumber \\
    &
    -(2+c_{13}+3c_2)
    (c_1^3+c_1c_3(-2+3c_3)+c_1^2(2+5c_3)
    -c_3^2(c_3-4c_4)
    +4c_1(1+c_3)c_4)\omega^3
    \Big)
    \nonumber \\
    &
    \times k_{a}\big(\bar{\nabla}_{a}\bar{u}_{b}
    -\bar{\nabla}_{b}\bar{u}_{a}\big)
    \overset{(0)}{N^{b}}
    \nonumber \\
    &
    -2 c_{123} {}^{(3)}k^2
    \Big(-(c_{13}^2-2v_{T}^{-2}c_4)
    \omega \bar{u}^{b}\bar{\nabla}_{b}\overset{(0)}{N_{a}}
    +2v_{T}^{-2}v_{V}^2c_{14}
    k_{b}\bar{\nabla}^{b}\overset{(0)}{N_{a}}\Big)
    k^{a}
    \nonumber \\
    &
    -2 c_{13} (2+c_{13}+3c_2)\omega
    \Big((c_{13}^2-2v_{T}^{-2}c_4)\omega^{2}
    \bar{u}^{b}\bar{\nabla}_{b}\overset{(0)}{N_{a}}
    \nonumber \\
    &
    -2v_{T}^{-2}v_{V}^2c_{14}\omega
    k_{b}\bar{\nabla}^{b}\overset{(0)}{N_{a}}\Big)
    k^{a}
    + \Big(
    -2v_{T}^{-2}v_{V}^2c_{14}c_{123}{}^{(3)}k^4
    \nonumber \\
    &
    +\big(
    -c_1^4 + c_3^3(2+3c_2+c_3)
    - c_1^3(2+3c_2+2c_3)
    \nonumber \\
    &
    -2(-1+c_3)c_{23}c_4
    + c_1^2(c_2(4-3c_3)-2(-3+c_{34}))
    \nonumber \\
    &
    + c_1(c_2(2+4c_3+3c_3^2-2c_4)
    +2(c_3(3+c_3+c_3^2-2c_4)+c_4))\big)
    {}^{(3)}k^2\omega^{2}
    \nonumber \\
    &
    -2v_{T}^{-2}c_{13}(2+c_{13}+3c_2)c_{14}\omega^{4}
    \Big)
    \bar{\gamma}_{ab}\bar{\nabla}^{b}\overset{(0)}{N^{a}}
    +\Big(
    -4v_{T}^{-2}v_{V}^{2}c_{14} c_{123}
    {}^{(3)}k^4\bar{u}^{b}\bar{\nabla}_{b}\bar{u}^{a}
    \nonumber \\
    &
    -(2+c_{13}+3c_2)c_{14}\omega^{3}
    \Big(-(c_{13}^2-2v_{T}^2c_4)\omega
    \bar{u}^{b}\bar{\nabla}_{b}\bar{u}^{a}
   -2v_{T}^{-2}v_{V}^{2}c_{14}
    k_{b}
    (\bar{\nabla}^{a}\bar{u}^{b}-\bar{\nabla}^{b}\bar{u}^{a})
    \Big)
    \nonumber \\
    &
    +{}^{(3)}k^2\omega
    \Big(-
    (2c_1^4 + 4c_3^2 -2c_3^3(2+c_3)
    + c_1^3(1+6c_2+4c_3)
    -6c_2c_4
    \nonumber \\
    &
    +(-2+c_3)c_3c_4
    +2(-1+c_3)c_4^2
    +2c_2c_3(2+c_3-3c_3^2+2c_4)
    \nonumber \\
    &
    +c_1^2(-6-2c_3+2c_2(-5+3c_3)+3c_4)
    \nonumber \\
    &
    +c_1(
    -c_3(2+c_3(7+4c_3))
    +4(-1+c_3)c_4+2c_4^2
    \nonumber \\
    &
    +c_2(2-2c_3(4+3c_3)+4c_4)
    )
    )\omega\bar{u}^{b}\bar{\nabla}_{b}\bar{u}^{a}
    \nonumber \\
    &
    -2v_{T}^{-2}v_{V}^{2}c_{14}(c_{14}+2c_2)
    k_{b}
    (\bar{\nabla}^{a}\bar{u}^{b}-\bar{\nabla}^{b}\bar{u}^{a})
    \Big)\Big)\overset{(0)}{N_{a}}
    \Big].    
\end{eqnarray}	

\subsection{Scalar modes}
\label{Scalar modes}
For scalar modes, the amplitude evolution equation is
\begin{eqnarray}
	\label{scalar part of the next-to-leading-order Scalar modes}
	&
	\frac{2i\big(2(c_{23}-c_4)+(-4c_2+c_{13}(c_{13}+3c_2))c_{14}\big)}
	{c_1+2c_3-c_4}\,\bar{u}_{a}\,\bar{\nabla}^{a}\overset{(0)}{L}
	+\frac{i\big(2(c_{23}-c_4)+(-4c_2+c_{13}(c_{13}+3c_2))c_{14}\big)}
	{c_1+2c_3-c_4}\,\bar{\nabla}_{a}\bar{u}^{a}\!\overset{(0)}{L}
	\nonumber \\
	&+\frac{i\,c_{123}(-2+c_{14})}
	{(c_1+2c_3-c_4)\omega}\,
	\bar{\nabla}_{a}k^{a}\!\overset{(0)}{L}
	+\frac{2i\,c_{123}(-2+c_{14})}
	{(c_1+2c_3-c_4)\omega}k^{a}\,\bar{\nabla}_{a}\overset{(0)}{L}
	+\frac{iv_{S}^2}
	{(c_1+2c_3-c_4)\omega}
	\;{}^{(3)}k_{a}\,\bar{u}_{b}\,\bar{\nabla}^{a}\bar{u}^{b}\,\overset{(0)}{L}
	\nonumber \\
	&\times
	\big(
	c_1^{3} - c_3^{2} - 3c_2c_3(-1+c_4) + 6c_4 - 5c_3c_4 - c_4^{2}
	+ c_1^{2}(-2 - 3c_2 + c_{34})
	\nonumber \\
	&
	- c_1\big(c_3 + 8c_4 - c_3c_4 + 3c_2(-1 + c_{34})\big)
	\big)
	+\frac{2i v_{S}^2}
	{(c_1+2c_3-c_4)\omega^{2}}
	\;{}^{(3)}k_{a}\,\bar{\nabla}^{a}\omega\,\overset{(0)}{L}
	\nonumber \\
	&\times
	\big(
	c_1^{5} - 2c_3 - 4c_3^{2} + 4c_4 + 3c_3c_4 - 2c_3^{3}c_4 - 2c_3c_4^{2}
	+ c_3^{3}c_4^{2}
	\nonumber \\
	&
	+ c_1^{4}(-2 + 4c_2 + 3c_3 + 2c_4)
	+ c_2^{2}(-2 + c_4)\big(2 + (-4 + 3c_3)c_4\big)
	\nonumber \\
	&
	+ c_1^{3}\big(-2 + 3c_2^{2} + 3c_3^{2} + 6c_3(-1 + c_4)
	- 2c_4 + c_4^{2}
	+ 4c_2(-3 + 2c_3 + 2c_4)\big)
	\nonumber \\
	&
	+ c_2\big(-2+11c_4 + 4c_3^{2}(-2 + c_4)c_4 - 2c_4^{2}
	+ c_3(-8 + 6c_4 - 4c_4^{2})\big)
	\nonumber \\
	&
	+ c_1^{2}\big(-1 + c_3^{3} + 6c_3^{2}(-1 + c_4) - 4c_4
	+ c_2^{2}(-10 + 3c_3 + 6c_4)
	\nonumber \\
	& + c_3(-2 - 6c_4 + 3c_4^{2})
	+ 4c_2(1 + c_3^{2} - 4c_4 + c_4^{2} + c_3(-5 + 4c_4))\big)
	\nonumber \\
	&
	+ c_1\big(2 + 2c_3^{3}(-1 + c_4) + 3c_4 + 3c_3^{2}(-2 + c_4)c_4
	- 2c_4^{2} - c_3(5 + 4c_4)
	\nonumber \\
	&
	+ c_2^{2}(10 + 6c_3(-1 + c_4) - 14c_4 + 3c_4^{2})
	\nonumber \\
	&
	+ c_2(3 + 8c_3^{2}(-1 + c_4) + 2c_4 - 4c_4^{2}
	+ c_3(6 - 24c_4 + 8c_4^{2})) )
	\big)
	\nonumber \\
	&
	-\frac{i}{(c_1+2c_3-c_4)\omega}\,\bar{u}_{a}\,\bar{\nabla}^{a}\omega\,
	\overset{(0)}{L}
	\Big(
	c_{1}^{3}
	- 2c_{3}
	- 2c_{4}
	+ 2c_{3}c_{4}
	+ c_{3}^{2}c_{4}
	+ c_{1}^{2}(2 + 3c_{2} + 2c_{3} + c_{4})
	\nonumber \\
	&+ c_{2}(-2 - 2c_{4} + 3c_{3}c_{4})
	+ c_{1}(-4 + 2c_{3} + c_{3}^{2} + 2c_{4}
	+ 2c_{3}c_{4} + c_{2}(-2 + 3c_{3} + 3c_{4}))
	\Big)
	\nonumber\\
	&
	-\frac{iv_{S}^2c_{14}}
	{c_{123}(c_1+2c_3-c_4)}
	\,\bar{u}_{a}\bar{u}_{b}\,\bar{\nabla}^{b}\bar{u}^{a}\;\overset{(0)}{L}
	\Big(
	2c_{1}^{3}
	+ 3c_{2}^{2}(-2 + c_{3})
	- 2c_{3}^{2}
	+ c_{3}^{3}
	+ c_{1}^{2}(-4 + 5c_{2} + 5c_{3})
	\nonumber \\
	&
	+ 2c_{4}
	- 3c_{3}c_{4}
	+ c_{2}(4c_{3}^{2} + c_{4} - 2c_{3}(4 + c_{4}))
	\nonumber \\
	&
	+ c_{1}(3c_{2}^{2} - 6c_{3} + 9c_{2}c_{3}
	+ 4c_{3}^{2} - 3c_{4} - 2c_{2}(5 + c_{4}))
	\Big)
	\nonumber \\
	&
	-\frac{2i\,c_{123}(-2+c_{14})}
	{(c_1+2c_3-c_4)\omega^{2}}\,
	{}^{(3)}k_{a}{}^{(3)}k_{b}\,\bar{\nabla}^{b}\bar{u}^{a}\,\overset{(0)}{L}
	-\frac{2i\,c_{123}(-2+c_{14})}
	{(c_1+2c_3-c_4)\omega}\,
	{}^{(3)}k_{a}\,\bar{u}_{b}\,\bar{\nabla}^{b}\bar{u}^{a}\,\overset{(0)}{L}
	=0.
\end{eqnarray}	

The mixed tensor mode amplitude associated with the leading-order scalar mode satisfies
\begin{eqnarray}
	\label{tensor part of the next-to-leading-order Scalar modes}
	&\epsilon\overset{(1)}{E_{fg}}
	=
	\frac{-i(
		\mathcal{A}\omega^{2}
		-\mathcal{B}\,k^2
		)}{(c_1+2c_3-c_4)\,\omega\,{}^{(3)}k^{4}\big(c_{13}\omega^{2}
		+ k^2\big)} \overset{(0)}{L}\,
	\Big[
	{}^{(3)}k_{f}{}^{(3)}k_{g}\,{}^{(3)}k^{2}\,\bar{\gamma}_{ab}\,\bar{\nabla}^{b}\bar{u}^{a}
	\nonumber \\
	&
	+ \,{}^{(3)}k_{a}\Big\{
	{}^{(3)}k_{g}\Big(
	{}^{(3)}k_{b}{}^{(3)}k_{f}\,\bar{\nabla}^{b}\bar{u}^{a}
	+{}^{(3)}k^{2}\Big(
	-\bar{\nabla}_{a}\bar{u}_{f}
	-
	\bar{u}_{b}\bar{u}_{f}\,\bar{\nabla}^{b}\bar{u}^{a}
	-\bar{\nabla}_{f}\bar{u}^{a}
	\Big)
	\Big)
	\nonumber \\
	&
	+{}^{(3)}k^{2}\Big(
	k_{b}\,\bar{\gamma}_{fg}\,\bar{\nabla}^{b}\bar{u}^{a}
	+ {}^{(3)}k_{f}\,\Big(-\bar{\nabla}_{a}\bar{u}_{g}
	-
	\bar{u}_{b}\,\bar{u}_{g}
	\bar{\nabla}^{b}\bar{u}^{a}
	-\bar{\nabla}_{g}\bar{u}^{a}
	\Big)
	\Big)
	\Big\}
	\nonumber \\
	&
	+{}^{(3)}k^{4}\Big\{
	\bar{u}_{a}\big(
	\bar{u}_{g}\,\bar{\nabla}_{a}\bar{u}_{f}
	+\bar{u}_{f}\,\bar{\nabla}_{a}\bar{u}_{g}
	\big)
	-\bar{\gamma}_{ab}\bar{\gamma}_{fg}\,\bar{\nabla}^{b}\bar{u}^{a}\,
	+ 
	\big(\bar{\nabla}_{f}\bar{u}_{g}+\bar{\nabla}_{g}\bar{u}_{f}\big)
	\Big\}
	\Big].
\end{eqnarray}	
Here,
\begin{eqnarray}
	\label{AB=}
	\mathcal{A}
	\!&\coloneqq &\!
	4c_1-6c_1^{2}+3c_1^{3}-4c_1c_3+6c_1^{2}c_3+2c_3^{2}
	+3c_1c_3^{2}+(-2+c_{13})(-2+3c_{13})c_4,
	\nonumber \\
	\mathcal{B}
	\!&\coloneqq&\!
	c_{13}(-2+c_{14}).
\end{eqnarray}	

Finally, the mixed vector modes satisfy
\begin{eqnarray}
	\label{vector part of the next-to-leading-order Scalar modes B}
	&\epsilon\overset{(1)}{B_{g}}
	=
	c_{13}\,\epsilon\overset{(1)}{N}_{g}
	-\frac{1}{(c_1+2c_3-c_4)\,{}^{(3)}k^{4}\,\omega^{2}}\;
	i\,{}^{(3)}k^{a}\,\overset{(0)}{L}
	\nonumber \\
	&
	\times
	\Big[
	-2\,{}^{(3)}k_{b}\,{}^{(3)}k_{g}\,
	\bar{\nabla}^{b}\bar{u}_{a}\,
	\Big(
	\mathcal{C}\omega^{2}
	- \mathcal{D}\,k^2
	\Big)
	\nonumber \\
	&
	+ {}^{(3)}k^{2}\Big(
	\mathcal{C}\bar{u}^{b}\bar{u}_{g}\,\omega^{2}\,\bar{\nabla}_{b}\bar{u}_{a}
	- \mathcal{D}\,{}^{(3)}k_{g}\,
	\Big(
	2\,\bar{\nabla}_{a}\omega - \bar{\nabla}_{a}\bar{u}^{b}k_{b}
	-2\bar{u}^{b}\,\bar{\nabla}_{b}k_{a}
	+k_{b}\,\bar{\nabla}^{b}\bar{u}_{a}
	\Big)
	\nonumber \\
	&
	+ \bar{\nabla}_{a}\bar{u}_{g}\,
	\Big(
	\mathcal{C}\omega^{2}
	- \mathcal{D}\,k^2
	\Big)
	+\mathcal{C}\omega^{2}\,\bar{\nabla}_{g}\bar{u}_{a}
	-\mathcal{D}\bar{u}^{b}\bar{u}_{g}\,\bar{\nabla}_{b}\bar{u}_{a}\,
	k^2
	-\mathcal{D}k^2\,\bar{\nabla}_{g}\bar{u}_{a}
	\Big)
	\Big]
	\nonumber \\
	&
	- \frac{i}{(c_1+2c_3-c_4)\,\omega^{2}}\mathcal{D}
	\overset{(0)}{L}\,
	\Big(
	k_{a}(\bar{\nabla}^{a}\bar{u}_{g}-\bar{\nabla}_{g}\bar{u}^{a})
	\nonumber \\
	&
	-2\,\bar{\nabla}_{g}\omega
	+ \bar{u}^{a}\bar{u}_{g}\,(-2\bar{\nabla}_{a}\omega - \bar{\nabla}_{a}\bar{u}^{b}k_{b}
	-2\bar{u}^{b}\bar{\nabla}_{b}k_{a})
	-2\,\bar{u}^{a}\bar{\nabla}_{g}k_{a}
	\Big),
	\\
	\label{vector part of the next-to-leading-order Scalar modes N}
	&\epsilon\overset{(1)}{N_{g}}
	=- \frac{i}{(c_1+2c_3-c_4)\,{}^{(3)}k^{4}\,\omega^{2}}\,
	\nonumber \\
	&\times\frac{1}
	{
	\big(
	-2c_1+c_1^2-c_3^2
	\big)\,k^2
	-\big(
	c_1^2+c_3^2-2c_4+2c_3c_4+2c_1c_{34}
	\big)\,
    \omega^{2}
	}
	\Big\{
	2\,{}^{(3)}k_{a}\,{}^{(3)}k_{b}\,{}^{(3)}k_{g}\,
	\overset{(0)}{L} \,\bar{\nabla}^{b}\bar{u}^{a}\,
	\nonumber \\
	&
	\times\Big[
	(c_1-c_3)k^2\,
	\Big(
	\mathcal{C}\,\omega^2
	-\mathcal{D}k^2
	\Big)
	- (c_{13}+2c_4)\,\omega^2\,
	\Big(
	\mathcal{C}\,\omega^2
	- \mathcal{D}k^2
	\Big)
	\Big]	
	\nonumber \\
	&-{}^{(3)}k^{a}{}^{(3)}k^2{}^{(3)}k_{g}
   \Big[\mathcal{E}\overset{(0)}{L}  \omega^2 \bar{\nabla}_a \bar{u}^b k_b
   	+\mathcal{F} \overset{(0)}{L} \omega^2 k_b \bar{\nabla}^b \bar{u}_a
   	\nonumber \\
   	&
   	+2c_{123}(-2+c_{14})\omega (\bar{\nabla}_a \overset{(0)}{L}) k^2
   	+4c_{123}(-2+c_{14}) \overset{(0)}{L} \omega \bar{\nabla}_b k_a k^b
   	\nonumber\\
   	& 
   	+ 2\big(c_1^2-c_3^2-c_{123}\big)(-2+c_{14}) \overset{(0)}{L} (\bar{\nabla}_a \omega) k^2
   \nonumber\\
   	& +(c_1-c_3)\mathcal{D} \overset{(0)}{L} \bar{\nabla}_{a} \bar{u}_b	k^b k^2
   	+ (-2c_1^2+c_1^3+2c_3^2) \overset{(0)}{L}\omega
   		k^2
   		\nonumber\\
   		& 
   		+ (c_1c_3^2-c_1^2c^4+c_3^2c_4) \overset{(0)}{L} k_b \bar{\nabla}^{b}\bar{u}_a k^2  
   		\nonumber\\
   			&+ \bar{u}^{b}
   			\Big(-2 \omega \, \bar{\nabla}_a \overset{(0)}{L} \, k_b
   			\Big[
   			-\mathcal{G}\omega
   			+ (c_{23} - c_4)(-2 + c_{14})\, \omega
   			\Big] 
   			\nonumber\\
   			& + \overset{(0)}{L} \Big(
   			2 (-2 + c_{14}) \omega\, k_c
   			\Big[
   			(2c_{23} + c_4) \bar{\nabla}_a \bar{u}^c k_b
   			+ (2 c_1 - c_4) k_b \bar{\nabla}^c \bar{u}_a
   			- c_4 \bar{\nabla}_b \bar{u}_a k^c
   			\Big]
   			\nonumber\\
   			& 
   			+ 2 \bar{\nabla}_b k_a
   			\Big[
   			\mathcal{G} \omega^2 
   			+ (c_1 - c_3)c_{13}(-2 + c_{14})
   			k^2
   			\Big]
   			\nonumber\\
   			& 
   			+ \bar{u}^c \Big(
   			2 \omega\,\bar{\nabla}_b \bar{u}_a\,k_c
   			\Big(
   			\mathcal{I}\omega
   			+ 4 c_4 (-2 + c_{14})\, \omega
   			\Big)
   			\nonumber\\
   			& 
   			+ (-2 + c_{14})k_b
   			\Big(
   			-\mathcal{J}
   			\bar{\nabla}_a \omega k_c
   			+ 4 (c_{23} - c_4) \omega \bar{\nabla}_c k_a 
   			\nonumber\\
   			& 
   			- c_{13}(c_{13} + 2 c_4)
   		k_c
   			\Big(
   			2\,\bar{u}^d\bar{\nabla}_d k_a
   			+k_d(\bar{\nabla}_a \bar{u}^d - \bar{\nabla}^d \bar{u}_a)
   			\Big)
   			\Big)
   			\Big)
   			\Big)
   			\Big)
   			 \Big]
   			\nonumber
   			\\
   				&+
   				{}^{(3)}k_{a}\,{}^{(3)}k^{2}\,\overset{(0)}{L}\,
   				\Big(
   				\bar{\nabla}^{a}\bar{u}_{g}\,
   				k_b
   				\Big(
   				-\Big((c_{1}-c_{3})\,
   		     	k^b
   				\big(
   				\mathcal{H}\omega^{2}
   				-\mathcal{D}\,
   				k^2
   				\big)\Big)
   				\nonumber \\ 
   				&
   				+(c_{13}+2c_{4})\,\bar{u}_{b}\bar{u}_{c}\, k^{c}
   				\big(
   				\mathcal{H}\omega^{2}
   				-\mathcal{D}
   				k^2
   				\big)
   				\Big)
   				-(c_{1}-c_{3})\,
   				k^2
   				\big(
   				\mathcal{H}\omega^{2}
   				-\mathcal{D}\,
   				k^2
   				\big)\,
   				\bar{\nabla}_{g}\bar{u}^{a}
   				\nonumber \\ 
   				&
   				+\bar{u}_{b}\,k_c
   				\Big[
   				\bar{u}_{g}\,\bar{\nabla}^{b}\bar{u}^{a}
   				\Big(
   				-\Big((c_{1}-c_{3})\,
   				k^{c}
   				\big(
   				\mathcal{H}\omega^{2}
   				-\mathcal{D}
   				k^2
   				\big)
   				\Big)
   				+(c_{13}+2c_{4})\,\bar{u}_{c}\bar{u}_{d}\,
   				k^{d}
   				\big(
   			    \mathcal{H}\omega^{2}
   				-\mathcal{D}\,
   				k^2
   				\big)
   				\Big)
   				\nonumber \\ 
   				&
   				+(c_{13}+2c_{4})\,
   				\bar{u}_{c}\,k_b
   				\Big(
   				\mathcal{H}\omega^{2}
   				-\mathcal{D}\,
   				k^2
   				\Big)
   				\bar{\nabla}_{g}\bar{u}^{a}
   				\Big]
   				\Big)
   				\nonumber
   		+4{}^{(3)}k^4c_{123}(-2+c_{14})\omega k^a\bar{\nabla}_{g}k_a \overset{(0)}{L}
   		\nonumber \\
   		& 
   		+{}^{(3)}k^{4}\,\bar{u}^{a}\Big(
   		\bar{u}_{g}\Big(
   		(4-2c_{14})
   		\Big(
   		\big(-c_{123}\,\omega\,\bar{\nabla}_{a}\overset{(0)}{L}
   		-(c_{1}^2-c_{3}^{2}-c_{123})\,\overset{(0)}{L}\,\bar{\nabla}_{a}\omega\big)
   		k_{b}
   		-2c_{123}\,\overset{(0)}{L}\,\omega\,\bar{\nabla}_{b}k_a
   		\Big)
   		k^b
   		\nonumber \\
   		& 
   		+\overset{(0)}{L}\,
   		\bar{\nabla}_{a}\bar{u}_{b}\,
   		k^b
   		\Big(
   		\mathcal{E}\omega^{2}
   		+(c_{1}-c_{3})\mathcal{D}
   		k^2
   		\Big)
   		\Big)
   		+\bar{u}_{b}\Big(
   		2\,\overset{(0)}{L}\,\omega\,
   		\bar{\nabla}_{a}\bar{u}_{g}\,
   		k^b
   		\Big(
   		\mathcal{I}\omega
   		+4c_{4}(-2+c_{14})\omega
   		\Big)
   		\nonumber \\
   		& 
   		+\bar{u}_{g}\Big(
   		2\omega\,\bar{\nabla}_{a}\overset{(0)}{L}\,
   		k_b
   		\big(
   		\mathcal{G}\omega
   		-(c_{23}-c_{4})(-2+c_{14})\omega
   		\big)
   		\nonumber \\
   		& 
   		+\overset{(0)}{L}\Big(
   		2\,\bar{\nabla}_{b}k_a
   		\big(
   	\mathcal{G}\omega^{2}
    +(c_{1}-c_{3})\mathcal{D}k^2
   		\big)
   		-( -2+c_{14})
   		k_bk_c
   		\Big(
   		-2(2c_{23}+c_{4})\omega\,\bar{\nabla}_{a}\bar{u}^{c}
   		\nonumber \\
   		&
   		+\mathcal{J}
   		\bar{u}^{c}\bar{\nabla}_{a}\omega
   		+c_{13}(c_{13}+2c_{4})
   		\bar{u}_{c}
   		\bar{\nabla}_{a}\bar{u}_{d}\,
   		k^d
   		\Big)
   		\Big)
   		\Big)
   		\Big)
   		\nonumber \\
   		& 
   		+2\overset{(0)}{L}\Big(
   		-c_{4}(-2+c_{14})\omega\,
   		\bar{\nabla}_{a}\bar{u}_{g}\,
   		k^2
   		+\big(
   		\mathcal{G}\omega^{2}
   		+(c_{1}-c_{3})\mathcal{D}
   		k^2
   		\big)
   		\bar{\nabla}_{g}k_a
   		\Big)
   		\Big)
   		\nonumber
   		\\
	&+{}^{(3)}k^4 
	k_a
	\Big[
	\overset{(0)}{L}\,
	\bar{\nabla}^{a}\bar{u}_{g}
	\Big(
	\mathcal{F}\omega^2 
	- (c_1 - c_3)\mathcal{D}
	k^2
	\Big)
	+2c_{123}(-2+c_{14}) \omega k^a\bar{\nabla}_{g}\overset{(0)}{L}
	+\mathcal{E} \overset{(0)}{L}\,\omega^2\,\bar{\nabla}_{g}\bar{u}^a
	\nonumber\\
	&
    +(c_1-c_3)\mathcal{D} \overset{(0)}{L}\,
	k^2\bar{\nabla}_{g}\bar{u}^a
	+ 2(c_1^2-c_3^2-c_{123})(-2+c_{14}) \overset{(0)}{L}\,
	k^a\bar{\nabla}_{g}\omega
	\nonumber\\
	&
	+ \bar{u}_a
	\Big\{
	-2\omega\Big[
	-\mathcal{G}\omega
	\bar{\nabla}_{g}\overset{(0)}{L}
	- 2 (c_{23} - c_4)(-2 + c_{14})
	\overset{(0)}{L}\,
	\bar{u}^b\big(
	\bar{u}^c\,\bar{u}_{g}\,\bar{\nabla}_{c}k_b
	+ \bar{\nabla}_{g}k_b
	\big)
	\Big]
	\nonumber\\
	&
	+ (-2 + c_{14})k_b
	\Big[
	2(2 c_1 - c_4)\overset{(0)}{L}\,\omega\,\bar{\nabla}^b\bar{u}_{g}
	+ 2(2c_{23}+c_4)\overset{(0)}{L}\,\omega\,\bar{\nabla}_{g}\bar{u}^b
	\nonumber\\
	&
	+ \bar{u}^b\Big(
	2(c_{23}-c_4)\omega\bar{\nabla}_{g}\overset{(0)}{L}
	+ c_{13}(c_{13}+2 c_4)\overset{(0)}{L}\,
	k_c(\bar{\nabla}^c\bar{u}_{g}
	- \bar{\nabla}_{g}\bar{u}^c)
		\nonumber\\
	&
	- \mathcal{J}
	\overset{(0)}{L}\bar{\nabla}_{g}\omega
	- 2c_{13}(c_{13}+2c_4)\overset{(0)}{L}
	\bar{u}^c\big(
	\bar{u}^d\,\bar{u}_{g}\,\bar{\nabla}_{d}k_c
	+ \bar{\nabla}_{g}k_c
	\big)
	\Big)
	\Big]
	\Big\}
	\Big]
	\Big].
\end{eqnarray}
Here,
\begin{eqnarray}
\label{C,D,E,F,G=}
\mathcal{C}
\!&\coloneqq&\!
3c_1^{3}+4c_4-8c_3c_4
+3c_1^{2}(-2+2c_3+c_4)
\nonumber \\
\!&+&\! 
c_3^{2}(2+3c_4)
+ c_1(4+3c_3^{2}-8c_4 + c_3(-4+6c_4)),
\nonumber \\
\mathcal{D}
\!&\coloneqq&\!
c_{13}(-2+c_{14}),
\nonumber \\
\mathcal{E}
\!&\coloneqq&\!
2\big(3c_1^3+2c_3^2(-1+c_4)+3(-1+c_2)c_3c_4
+c_4(-3c_2+c_4)
+c_1^2(-2+3c_2+5c_3+3c_4)
\nonumber \\
\!&+&\! 
c_1(c_3(-1+2c_3)+5(-1+c_3)c_4+3c_2(-1+c_{34}))
\big),
\nonumber \\
\mathcal{F}
\!&\coloneqq&\!
2\big(
c_1^3+c_1^2(-2+3c_3+c_4)
+c_1c_3(-2+2c_3+3c_4)
+2c_3(-2c_4+c_3(2+c_4))
\big),
\nonumber \\
\mathcal{G}
\!&\coloneqq&\!
\big(-3 c_2 + c_3 + c_{13}(c_{13} + 3 c_2 ) - c_4\big)c_{14},
\nonumber \\
\mathcal{H}
\!&\coloneqq&\!
4c_{1}-6c_{1}^{2}+3c_{1}^{3}-4c_{1}c_{3}+6c_{1}^{2}c_{3}
+2c_{3}^{2}+3c_{1}c_{3}^{2}+(-2+c_{13})(-2+3c_{13})c_{4},
\nonumber \\
\mathcal{I}
\!&\coloneqq&\!
2 c_3(c_1^2 + (2 + c_1)c_3)
+ (-c_1^2 + 2(-1 + c_3)c_3 + c_1(2 + c_3)) c_4 
- c_{13}c_4^2,
\nonumber \\
\mathcal{J}
\!&\coloneqq&\!
2(c_{23}-c_{4}+c_{13}(c_{13}+2c_{4})).
\end{eqnarray}

\end{document}